\documentclass[a4paper,twocolumn,11pt, accepted=2023-04-29]{quantumarticle}
\pdfoutput=1

\usepackage[table]{xcolor}
\usepackage[utf8]{inputenc}
\usepackage[english]{babel}
\usepackage[T1]{fontenc}
\usepackage{amsmath}
\usepackage{hyperref}

\usepackage{tikz}
\usepackage{lipsum}

\usepackage{bm}
\usepackage{braket}
\usepackage{listings}
\usepackage{url}
\usepackage{amsfonts}
\usepackage{floatrow}
\usepackage{makecell}
\usepackage{pbox}
\usepackage{array}

\usepackage[defaultmono]{droidsansmono}
\usepackage[T1]{fontenc}

\definecolor{codegreen}{HTML}{007662}
\definecolor{codeorange}{HTML}{ff7f32}
\definecolor{codepurple}{HTML}{70309f}
\definecolor{codegray}{rgb}{0.5,0.5,0.5}
\definecolor{backcolour}{HTML}{ebeff1}

\lstdefinestyle{codeStyle}{
    backgroundcolor=\color{backcolour},   
    commentstyle=\color{codegreen},
    keywordstyle=\color{codeorange},
    numberstyle=\tiny\color{codegray},
    stringstyle=\color{codepurple},
    basicstyle=\fontfamily{droidsansmono}\scriptsize,
    breakatwhitespace=false,         
    breaklines=true,                 
    captionpos=b,                    
    keepspaces=true,                 
    numbers=left,                    
    numbersep=5pt,                  
    showspaces=false,                
    showstringspaces=false,
    showtabs=false,                  
    tabsize=2,
    frame=tlbr,
    framesep=2pt,
    framerule=0pt,
}

\lstdefinestyle{outputStyle}{
    backgroundcolor=\color{white},   
    commentstyle=\color{codepurple},
    keywordstyle=\color{codeorange},
    numberstyle=\tiny\color{codegray},
    stringstyle=\color{codegreen},
    basicstyle=\fontfamily{droidsansmono}\linespread{1.1}\scriptsize,
    breakatwhitespace=false,         
    breaklines=true,                 
    captionpos=b,                    
    keepspaces=true,                 
    numbers=none,                    
    numbersep=5pt,                  
    showspaces=false,                
    showstringspaces=false,
    showtabs=false,                  
    tabsize=2
}

\newcommand{\code}[1]{\mbox{\fontfamily{droidsansmono}\selectfont\footnotesize #1}}
\newcommand{\str}[1]{\mbox{\fontfamily{droidsansmono}\selectfont\footnotesize\color{codepurple} #1}}
\newcommand{\link}[1]{\href{https://#1}{\fontfamily{droidsansmono}\selectfont\scriptsize #1}}
\newcommand{\ext}{\text{ext}}
\newcommand{\ha}{\text{ha}}
\newcommand{\ch}{\text{ch}}
\newcommand{\nn}{n_\text{N}}
\newcommand{\nh}{n_\text{H}}
\newcommand{\nc}{n_\text{C}}
\newcommand{\nl}{n_\text{L}}
\newcommand{\qp}{\text{qp}}
\newcommand{\ie}{\textit{i.e.}}
\def\be{\begin{equation}}
\def\ee{\end{equation}}
\def\bea{\begin{eqnarray}}
\def\eea{\end{eqnarray}}

\begin{document}
\title{Analysis of arbitrary superconducting quantum circuits accompanied by a Python package: SQcircuit}
\author{Taha Rajabzadeh}
\email{tahar@stanford.edu}
\affiliation{Department of Electrical Engineering, Stanford University, Stanford, CA 94305 USA}
\author{Zhaoyou Wang}
\affiliation{E. L. Ginzton Laboratory and the Department of Applied Physics, Stanford University, Stanford, CA 94305 USA}
\author{Nathan Lee}
\affiliation{E. L. Ginzton Laboratory and the Department of Applied Physics, Stanford University, Stanford, CA 94305 USA}
\author{Takuma Makihara}
\affiliation{E. L. Ginzton Laboratory and the Department of Applied Physics, Stanford University, Stanford, CA 94305 USA}
\author{Yudan Guo}
\affiliation{E. L. Ginzton Laboratory and the Department of Applied Physics, Stanford University, Stanford, CA 94305 USA}
\author{Amir H. Safavi-Naeini}
\affiliation{E. L. Ginzton Laboratory and the Department of Applied Physics, Stanford University, Stanford, CA 94305 USA}
\email{safavi@stanford.edu}
\begin{abstract}
Superconducting quantum circuits are a promising hardware platform for realizing a fault-tolerant quantum computer. Accelerating progress in this field of research demands general approaches and computational tools to analyze and design more complex superconducting circuits. We develop a framework to systematically construct a superconducting quantum circuit's quantized Hamiltonian from its physical description. As is often the case with quantum descriptions of multicoordinate systems, the complexity rises rapidly with the number of variables. Therefore, we introduce a set of coordinate transformations with which we can find bases to diagonalize the Hamiltonian efficiently. Furthermore, we broaden our framework's scope to calculate the circuit's key properties required for optimizing and discovering novel qubits. We implement the methods described in this work in an open-source Python package \code{SQcircuit}. In this manuscript, we introduce the reader to the  \code{SQcircuit} environment and functionality. We show through a series of examples how to analyze a number of interesting quantum circuits and obtain features such as the spectrum, coherence times, transition matrix elements, coupling operators, and the phase coordinate representation of eigenfunctions.
\end{abstract}
\newpage
\renewcommand{\baselinestretch}{0.90}\normalsize
\tableofcontents
\renewcommand{\baselinestretch}{1}\normalsize
\section{Introduction}
\begin{figure*}[t]
  \centering
  \includegraphics[scale=0.6]{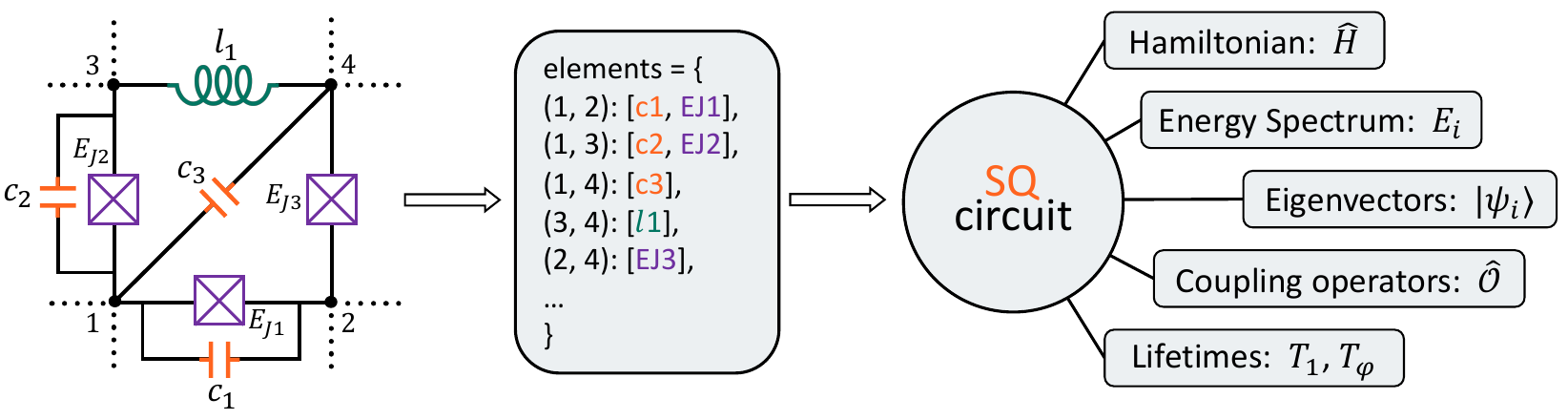}
  \caption{The general overview of \code{SQcircuit}. By defining the net-list of an arbitrary circuit in \code{SQcircuit} environment, users can compute and visualize quantities of interest such as energy spectra, coupling operators, and etc, through highly-abstracted routines.}
  \label{fig:intro}
\end{figure*}

Two of the important lines of research in superconducting quantum circuits that run parallel to the current quest to realize large-scale fault-tolerant quantum computers are (1) exploring the spaces of novel circuitry that more robustly encode quantum information~\cite{Gyenis2021, Krantz2019, Wang2021, rajabzadeh2020photonics, lee2023strong}; and (2) coupling superconducting qubits to other quantum systems to form hybrid devices~\cite{Xiang2013} that implement new memory, sensing, and communications functions. Both directions are characterized by increasing complexities in the circuit layout and demand a general approach toward analyzing larger and more complex circuits. To study the quantum behavior of a superconducting quantum circuit, we need to diagonalize the Hamiltonian of the circuit, derived from the circuit quantization method~\cite{Vool2017,Menke2021,Smith2016,kerman2020}.
However, the computational cost for diagonalization increases exponentially with the circuit size, and a proper basis choice is necessary for alleviating this problem when solving larger circuits. Existing numerical packages have made significant and valuable contributions in this direction~\cite{Gely2019,Groszkowski2021,Aumann2021}. We build on these earlier efforts by directly addressing the dual challenges of scalability and generality. We move beyond restrictions on the anharmonicity or types of qubits, and realize automatic constructions that enable more efficient diagonalization of larger circuits.

In this work, we develop a systematic framework that constructs the Hamiltonian for an arbitrary superconducting quantum circuit.
We pay close attention to proper accounting for time-dependent external fluxes~\cite{You2019, Riwar2021} and represent the Hamiltonian in a basis that requires fewer computational resources for diagonalization.
Additionally, we describe how to calculate properties essential to designing new superconducting qubits, such as coherence times in the presence of various noise processes, robustness to fabrication imperfections, and coupling to external control lines. We implement these approaches and provide an open-source Python-based package \code{SQcircuit}. Relying on NumPy, SciPy, and QuTiP~\cite{Johansson2011} packages, \code{SQcircuit} is a realization of the framework in this paper and provides a general approach for analyzing arbitrary circuits, complementing and extending the functionalities of the aforementioned packages~\cite{Groszkowski2021,Gely2019,Aumann2021}. Users can define arbitrary circuits and, through highly-abstracted routines, conveniently compute and visualize quantities of interest such as energy spectra, eigenvectors, and coherence properties (Fig.\ref{fig:intro}). We expect the package to significantly accelerate the analysis and design processes of complex superconducting circuits.

The remainder of the manuscript is structured as follows: Sec.~\ref{sec:generalHamil} reviews the general approach to obtaining the Hamiltonian for an arbitrary superconducting circuit. In Sec.~\ref{sec:transCoord}, we derive a proper transformation of coordinates to find an appropriate basis for efficiently diagonalizing the Hamiltonian of any superconducting circuit. In Sec.~\ref{sec:SQcircuitIntro}, we walk the reader through the SQcircuit library, and use the $0-\pi$ qubit~\cite{Brooks2013,Groszkowski2017,Gyenis2021Exp} as an example to demonstrate how to define a circuit in \code{SQcircuit} and obtain basic circuit properties such as eigenenergies and eigenvectors. Next, in Sec.~\ref{sec:features}, we expand the developed theory in previous sections to extract other circuit properties such as matrix elements, lifetimes, and so on, along with \code{SQcircuit} functionalities to obtain those features. We conclude by presenting an outlook in Sec.~\ref{sec:concolusion}.

\section{Circuit Hamiltonian}
\label{sec:generalHamil}

Our description of a quantum circuit follows previous studies~\cite{Yurke1984, Vool2017, Burkard2004}. We start with a circuit consisting of capacitors and inductors as branches. Applying Faraday's law around loops in the circuit leads to constraints that impose relationships between different branch fluxes and reduce the number of independent variables. A systematic approach to obtaining the independent degrees of freedom of the system is to define node fluxes by choosing a minimum spanning tree in the graph and assigning a flux variable $\phi_i$ at every node $i$~\cite{Vool2017}. Faraday's law then leads to a number of relations between the branch fluxes $\phi_{b,k}$ and node fluxes $\phi_i$ (subscript $b$ denotes branch variables), usually of the form  $\phi_{b,k} =\phi_i - \phi_j + \phi_{\text{ext},b,k}$, where nodes $i,j$ are the endpoints of $k$th branch. Here, $\phi_{\text{ext},b,k}$ is not a dynamical variable but a constraint imposed on branches that are not part of the minimum spanning tree. For these branches, the value of $\phi_{\text{ext},b,k}$ is found by considering the external flux in the loop formed by the branch (we refer to external fluxes of the $l$th loop by $\phi_{\text{ext},l}$ which is different from value of the external flux at the branch $\phi_{\text{ext},b,k}$). In a circuit with $\nn+1$ nodes and $\nl$ inductive loops, we have only $\nn$ independent variables, and we rewrite the relation between $k$th branch flux variable between nodes $i,j$ in terms of vectorized node fluxes $\bm{\Phi}= [\phi_1,...,\phi_{\nn}]$ (we have set the ground node $i=0$ to a fixed potential $\phi_0 = 0$) and external fluxes $\bm \Phi_{\text{ext}}=[\phi_{\text{ext},1},\dots,\phi_{\text{ext},\nl}]$ as:
\begin{equation}\label{eq:branchFlux}
    \phi_{b,k} = \bm{w}^T_k \bm \Phi + \bm{b}^T_k \bm \Phi_{\text{ext}},
\end{equation}
where $\bm{w}_k$ is the $\nn$ dimensional vector with $\delta_{im} - \delta_{mj}$ for element $m\in\{1,2,\cdots,\nn\}$ and $\bm{b}_k$ is the $\nl$ dimensional vector that its $l$th element is one if $k$th branch is a closure branch for loop $l$ otherwise it is zero.

The node fluxes take the role of position variables. There is an associated ``potential'' energy due to every inductive element. Every branch with an inductive element has an energy which is an instantaneous function of the branch flux, \ie, $U_{b,k} = U_{b,k}(\phi_{b,k})$.
We are primarily concerned with two types of inductors, linear inductors with inductance $l$ where $U_L(\phi_{b,k}) = \phi_{b,k}^2/2l $ and Josephson junctions  $U_J(\phi_{b,k}) = -E_J \cos(2\pi\phi_{b,k}/\Phi_0)$, where $\Phi_0=h/2e$ is the magnetic flux quantum and $E_J$ is the Josephson energy. 

For branches which are capacitors, the time-derivative of the branch flux is the voltage drop across the branch $V_{b,k}=\dot \phi_{b,k}$; therefore, we assign a ``kinetic'' energy $T(\dot \phi_{b,k})=c \dot \phi_{b,k}^2/2$ to any capacitor branch $k$ with a capacitance $c$.


\subsection{Purely Harmonic Circuits}

For the case of purely harmonic circuits, \ie, circuits containing only linear inductors and capacitors, the analysis is significantly simplified. As justified below, we need only to consider the inductances and capacitances between each pair of nodes $i$ and $j$ given by $l_{ij}$ and $c_{ij}$, respectively. Constant external fluxes can be absorbed into the flux variables -- the linearity of the circuit means that these shifts do not affect the dynamics. The total potential energy can then always be written as
\bea
U = \frac{1}{2}\sum_{ij} \frac{(\phi_i - \phi_j)^2}{l_{ij}},
\eea
while the total kinetic energy is given by
\bea
T = \frac{1}{2}\sum_{ij} c_{ij} {(\dot\phi_i - \dot\phi_j)^2},
\eea
where above summations are over $i,j=1,..,\nn$. These equations can be conveniently expressed in terms of vectors $\bm{\Phi}$ and $\dot{\bm{\Phi}}$, and by defining capacitance $\bm{C}$ and inverse inductance (susceptance) $\bm{L}^{*}$ matrices such that
\bea{}
U(\bm{\Phi}) &=& \frac{1}{2}\bm{\Phi}{}^T \bm{L}^{*} \bm{\Phi}{}, \nonumber\\
T(\dot{\bm{\Phi}}) &=& \frac{1}{2}\dot{\bm{\Phi}{}}^T \bm{C}  \dot{\bm{\Phi}{}}.
\eea{}
The relation between capacitance (susceptance) matrix elements and the capacitances (susceptances) of the branches is then
\begin{equation*}
C_{ij} = [\bm{C}]_{ij}:
 \begin{cases}
    C_{ii} = c_{i0} + \sum_{k\neq i}c_{ik}&i=j\\
    C_{ij} = -c_{ij}&i\neq j
 \end{cases}
\end{equation*}
and
\begin{equation*}
\frac{1}{L_{ij}} = [\bm{L}^{*}]_{ij}:
 \begin{cases}
    \frac{1}{L_{ii}} = \frac{1}{l_{i0}} + \sum_{k\neq i}\frac{1}{l_{ik}}&i=j\\
   \frac{1}{L_{ij}} = -\frac{1}{l_{ij}}&i\neq j
 \end{cases}
\end{equation*}
. The Lagrangian $\mathcal L(\bm{\Phi},\dot{\bm{\Phi}}) =T(\dot{\bm{\Phi}})-U(\bm{\Phi})$ gives us the dynamics of the system. The Euler-Lagrange equation generated by this Lagrangian for each variable $i$ is effectively Kirchhoff's current law for that node. In contrast, Kirchhoff's voltage law was automatically satisfied  when we applied the constraints from Faraday's law on the branch fluxes. 

As a prelude to obtaining a quantum theory, we find the Hamiltonian. For this, we first calculate the conjugate momenta or charge variables for each node flux $\phi_i$. The equation in vector form is given succinctly as the gradient
\bea{}
\bm{Q}{} &\equiv& \frac{\partial L}{\partial \dot{\bm{\Phi}{}}} = \bm{C} \dot{\bm{\Phi}{}}.
\eea{}
Assuming that the capacitance matrix is invertible, we obtain the circuit Hamiltonian $H$ through the Legendre transformation
\bea{}
H = \frac{1}{2} \bm{\Phi}{}^T \bm{L}^{*} \bm{\Phi}{} + \frac{1}{2} \bm{Q}{}^T \bm{C}^{-1} \bm{Q}{}.
\eea{}

\subsection{Circuits with Nonlinear Inductors}

The presence of nonlinear inductors, \ie, inductors where $U(\phi_b)$ is not quadratic, allows external fluxes to change the dynamics of circuits fundamentally. The simplest example is the superconducting quantum interference device (SQUID) which contains two Josephson junctions in parallel, forming a loop.

We first express the potential energy of the circuit in terms of the branch fluxes across all inductive elements in $\mathcal{S}_L \cup \mathcal{S}_J$, set of branches consisting of linear inductors $ \mathcal{S}_L $ and Josephson junctions $ \mathcal{S}_J $:
\begin{equation}
    U = \sum_{k\in \mathcal{S}_L \cup \mathcal{S}_J} U_k(\phi_{b,k}).
\end{equation}
But not all branch fluxes are independent variables. 
Using Eq.~\ref{eq:branchFlux}, we write the potential energy in terms of node flux variables $\bm \Phi$, and the external fluxes $\bm \Phi_{\text{ext}} = \Phi_0 \bm \varphi_{\text{ext}}/2\pi$:
\begin{equation*}
    \begin{split}
            U({\bm\Phi}) = &\frac{1}{2} {\bm\Phi}^T \bm{L}^{*} {\bm\Phi} +\sum_{k\in \mathcal{S}_L}\frac{1}{l_k}\left(\frac{\Phi_0}{2\pi}\right)\bm{w}^T_k{\bm{\Phi}}(\bm{b}_k^T\bm{\varphi}_{\text{ext}}) \\
            &- \sum_{k\in \mathcal{S}_J} E_{J_k} \cos \left(\frac{2\pi}{\Phi_0}\bm{w}^T_k{\bm{\Phi}}+\bm{b}_k^T\bm{\varphi}_{\text{ext}}\right)
    \end{split}
\end{equation*}

The kinetic part of the energy is given by summing over the capacitive energy of every capacitor:
\begin{equation*}
    T = \frac{1}{2} \sum_{k\in S_\mathcal{C}} c_k \dot \phi_{b,k}^2.
\end{equation*}
In the case where the external fluxes are time-independent ($\dot{\bm {\Phi}}_{\text{ext}}=0$), we find that 
\begin{equation*}
    \dot \phi_{b,k} =  \bm{w}^T_k \dot{\bm{\Phi}},
\end{equation*}
and 
\begin{equation*}
    T(\dot{\bm{\Phi}{}}) = \frac{1}{2}\dot{\bm{\Phi}{}}^T \bm{C}  \dot{\bm{\Phi}{}}.
\end{equation*}
as in the harmonic case. 

As noted recently~\cite{You2019, Riwar2021}, a time-dependent external flux poses a challenge to this description as we end up with terms $\dot \phi_i$ in the Lagrangian that depend on how we assign fluxes to branches. 
Since time-dependent flux is essential in contexts such as flux-noise-induced dephasing~\cite{You2019} and Floquet engineering~\cite{Huang2020}, 
it is crucial for \code{SQcircuit} to treat these cases correctly, and we provide details of our approach in Appendix~\ref{app:timeDependentFlux}.

From the Lagrangian description, we can again obtain the conjugate momenta, and through a Legendre transformation, the Hamiltonian, which is now given by:
\begin{equation}\label{eq:circuitH}
    \begin{split}
    {{H}} = &\frac{1}{2} {{\bm{Q}}}^T {\bm{C}}^{-1}{{\bm{Q}}} + \frac{1}{2} {{\bm\Phi}}^T {\bm{L}}^{*} {{\bm\Phi}}\\
    &+\sum_{k\in \mathcal{S}_L}\left(\frac{\Phi_0}{2\pi}\frac{\bm{b}_k^T\bm{\varphi}_{\text{ext}}}{l_k}\right){\bm{w}}^T_k{{\bm{\Phi}}}\\
    &-\sum_{k\in \mathcal{S}_J} E_{J_k} \cos \left(\frac{2\pi}{\Phi_0}{\bm{w}}^T_k{{\bm{\Phi}}}+\bm{b}_k^T\bm{\varphi}_{\text{ext}}\right).
    \end{split}
\end{equation}
Note that in the absence of Josephson junctions, we can absorb the external fluxes into $\bm \Phi$ and recover the Hamiltonian from the linear analysis. In Appendix \ref{app:exampleCircuit}, we show in detail how to construct the Hamiltonian for an example circuit.

Finally, to quantize our circuit, we promote every node flux and its conjugate momentum to operators ($\phi_i \rightarrow \hat \phi_i$ and so on...) with the canonically conjugate operators satisfying the commutation relation
\begin{equation*}
    [\hat{{\phi}}_m, \hat{{Q}}_n] = \delta_{mn}i\hbar.
\end{equation*}
In matrix form, the commutators are expressed compactly as
\begin{equation*}
    \hat{{\bm{\Phi}}} \hat{{\bm{Q}}}^T - \hat{{\bm{Q}}} \hat{{\bm{\Phi}}}^T = i\hbar \bm{1}, 
\end{equation*}
where $\bm{1}$ is the identity matrix.
\section{Coordinate Transformations and Hamiltonian Diagonalization}
\label{sec:transCoord}
A major challenge in performing a full quantum analysis of a superconducting circuit is that, in general, increasing the size of the circuit leads to an exponential increase in  the numerical resources needed for simulation.  Therefore, finding the coordinate transformations in which the Hamiltonian is sparse and falls off quickly from the diagonal is essential. Although we use exact diagonalization to calculate the eigenvalues, our method paves the way for approximate methods such as tensor networks \cite{di2021efficient} for large Hilbert spaces. To meet these criteria, we are free to transform the Hamiltonian of Eq.~\ref{eq:circuitH} by performing a canonical transformation of charge and flux operators:
\begin{align*}
    \hat{\tilde{\bm{Q}}} = \bm{R}^{-1}\hat{\bm{Q}},\\
    \hat{\tilde{\bm{\Phi}}} = \bm{S}^{-1}\hat{\bm{\Phi}},
\end{align*}
where $\bm{R}$ and $\bm{S}$ are $\nn\times \nn$ real invertible matrices. For convenience, we impose the requirement that the transformation is canonical, \ie, the transformed charge and flux operators satisfy the relations $[\hat{\tilde{\phi}}_m, \hat{\tilde{Q}}_n] = \delta_{mn}i\hbar$. This leads to the constraint
\begin{equation}
     \label{eq:transCriteria} \bm{S}^T = \bm{R}^{-1}.
\end{equation}
The transformed Hamiltonian has the following form:
\begin{equation}\label{eq:transfromedH}
    \begin{split}
    \hat{\tilde{H}} = &\frac{1}{2} \hat{\tilde{\bm{Q}}}^T \tilde{\bm{C}}^{-1}\hat{\tilde{\bm{Q}}} + \frac{1}{2} \hat{\tilde{\bm\Phi}}^T \tilde{\bm{L}}^{*} \hat{\tilde{\bm\Phi}}\\
    &+\sum_{k\in \mathcal{S}_L}\left(\frac{\Phi_0}{2\pi}\frac{\bm{b}_k^T\bm{\varphi}_{\text{ext}}}{l_k}\right)\tilde{\bm{w}}^T_k\hat{\tilde{\bm{\Phi}}}\\
    &-\sum_{k\in \mathcal{S}_J} E_{J_k} \cos \left(\frac{2\pi}{\Phi_0}\tilde{\bm{w}}^T_k\hat{\tilde{\bm{\Phi}}}+\bm{b}_k^T\bm{\varphi}_{\text{ext}}\right),
    \end{split}
\end{equation}
where the transformed inverse capacitance and susceptance matrices are:
\begin{align*}
    \tilde{\bm{C}}^{-1} = \bm{R}^T\bm{C}^{-1}\bm{R},\\
    \tilde{\bm{L}}^{*} = \bm{S}^T\bm{L}^{*}\bm{S},
\end{align*}
and the vector assigning fluxes to junctions is given by
\begin{equation*}
    \tilde{\bm{w}}_k^T = \bm{w}_k^T\bm{S}.
\end{equation*}
As explained in Appendix \ref{app:coordinateTransformation}, we can always find $\bm{S}$ and $\bm{R}$ such that $\tilde{\bm{L}}^{*}$ and $\tilde{\bm{C}}$ have the block diagonal form of:
\begin{align}
    \label{eq:transformedC}
    \tilde{\bm{C}}=\begin{bmatrix}
    \bm{C}^\ha  &   \bm{0}\\
    \bm{0}   &   \bm{C}^\ch
    \end{bmatrix},\\
    \label{eq:transformedL}
    \tilde{\bm{L}}^{*}=\begin{bmatrix}
    \bm{L}^{*^\ha}   &   \bm{0}\\
    \bm{0}   &   \bm{0}
    \end{bmatrix},
\end{align}
where the ``harmonic'' parts $\bm{C}^{\text{ha}}$ and $\bm{L}^{*^{\text{ha}}}$ are $\nh{\times}\nh$ invertible diagonal matrices, while the ``charge'' part $\bm{C}^{\text{ch}}$ is a $\nc{\times}\nc$ symmetric matrix (not necessarily diagonal). The corresponding transformed charge and flux operators are also divided into two parts:
\begin{align*}
    \begin{split}
        \hat{\tilde{\bm{Q}}}^T=&[\hat{\bm{Q}}^{\text{ha}^T}|\hat{\bm{Q}}^{\text{ch}^T}]\\
        =&[\hat{Q}_1^{\text{ha}}, \dots, \hat{Q}_{\nh}^{\text{ha}}| \hat{Q}_1^{\text{ch}}, \dots, \hat{Q}^{ch}_{\nc}],
    \end{split}\\
        \begin{split}
        \hat{\tilde{\bm{\Phi}}}^T=&[\hat{\bm{\Phi}}^{\text{ha}^T}|\hat{\bm{\Phi}}^{\text{ch}^T}]\\
        =&[\hat{\phi}_1^{\text{ha}}, \dots, \hat{\phi}_{\nh}^{\text{ha}}| \hat{\phi}_1^{\text{ch}}, \dots, \hat{\phi}_{\nc}^{\text{ch}}].
    \end{split}
\end{align*}
We similarly divide the $\tilde{\bm{w}}_k^T$ into to two parts:
\begin{equation*}
    \tilde{\bm{w}}_k^T= [\bm{w}^{\text{ha}^T}_k|\bm{w}^{\text{ch}^T}_k] 
\end{equation*}

The transformation that we described above expresses the system's dynamics as that of (1) $\nh$ uncoupled harmonic oscillators, with capacitances and inverse inductances given by the diagonals of $\bm{C}^\text{ha}$ and $\bm{L}^{*^{\text{ha}}}$, (2) $\nc$ superconducting islands with no conducting paths connecting them and charging energy given by the capacitance matrix $\bm{C}^\text{ch}$, and (3) the flux drop across several junctions as weighted by the vector $\tilde{\bm{w}}_k^T$. 
Next we will quantize the harmonic (1) and charge (2) modes separately, and eventually diagonalize and find the spectrum of the full circuit.

\paragraph{Harmonic modes:} Ignoring the junctions momentarily, the part of the Hamiltonian acting on the first subspace is given by $\frac{1}{2}\hat{\bm{Q}}^{\text{ha}^T}(\bm{C}^{\text{ha}})^{-1}\hat{\bm{Q}}^{\text{ha}} + \frac{1}{2}\hat{\bm{\Phi}}^{\text{ha}^T}\bm{L}^{*^{\text{ha}}}\hat{\bm{\Phi}}^{\text{ha}}$. Since $\bm{C}^{\text{ha}}$ and $\bm{L}^{*^{\text{ha}}}$ are diagonal matrices, we have $\nh$ uncoupled harmonic oscillators with charge and flux variables $\hat{Q}^\text{ha}_i$ and $\hat{\phi}^\text{ha}_i$. Our diagonalization has effectively found the normal mode decomposition of the system. We associate creation and annihilation operators $\hat{a}_i^\dagger$ and $\hat{a}_i$ with each of the harmonic oscillators. The charge and flux operators are then
\begin{align}
    \label{eq:chargeHarmonic}
    \hat{Q}^\text{ha}_i=i\sqrt{\frac{\hbar}{2Z_i}}(\hat a_i^\dag-\hat a_i,)\\
    \label{eq:fluxHarmonic}
    \hat{\phi}^\text{ha}_i=\sqrt{\frac{\hbar Z_i}{2}}(\hat a_i^\dag+\hat a_i),
\end{align}
where $Z_i$ and $\omega_i$ are the mode impedance and angular frequency respectively, and given by:
\begin{align}
    \label{eq:impedanceHarmonic}
    Z_i = \sqrt{1/(\bm{L}^{*^{\text{ha}}}_{i,i}\bm{C}^{\text{ha}}_{i,i})},\\
    \label{eq:angularFreq}
    \omega_i=\sqrt{\bm{L}^{*^{\text{ha}}}_{i,i}/\bm{C}^{\text{ha}}_{i,i}}.
\end{align}
\paragraph{Charge modes:} The remaining subspace has a kinetic energy term $\frac{1}{2}\hat{\bm{Q}}^{\text{ch}^T}(\bm{C}^{\text{ch}})^{-1}\hat{\bm{Q}}^{\text{ch}}$, and a potential energy that is only due to junctions. Consider the case where there is only a single charge mode, then the junction energy will have the form $\cos(\alpha \hat \phi_1^{\text{ch}} + \hat{O})$, where $\hat{O}$ is an operator that represents the rest of the terms in the cosine. Since we can write the cosine  as
\begin{equation*}
  \cos(\alpha \hat  \phi_1^{\text{ch}} + \hat{O}) = \frac{e^{i\alpha \hat  \phi_1^{\text{ch}}}e^{i\hat O} + e^{-i\alpha  \hat \phi_1^{\text{ch}}}e^{-i\hat O}}{2},
\end{equation*}
and $e^{i\alpha \hat  \phi_1^{\text{ch}}}$ effects a translation in the charge basis where $e^{\pm i\alpha \hat  \phi_1^{\text{ch}}}|\hat Q_1^{\text{ch}} = Q_1\rangle = |\hat Q_1^{\text{ch}} = Q_1 \pm \hbar\alpha\rangle$, we see that only a discrete number of charge eigenstates $|\cdots n_1 \cdots \rangle \equiv |\cdots (Q_1 + n_1 \hbar\alpha) \cdots \rangle$ for integer $n_1$ need to be considered. Moreover, we are free to choose the coefficient $\alpha$ with an appropriate transformation of the basis, \ie, a constant scaling of $\bm R$ and $\bm S$ which keeps the transformation canonical. This constant scaling is effectively equivalent to choosing the units of $\hat Q$ and $\hat \phi$ while keeping the commutation relation the same. A physical meaningful value for $\alpha$ is $2\pi/\Phi_0$. This choice means that Josephson tunneling changes the charge state by $2e$, and thus couples all charge states of the from $|n_1\rangle \equiv |\hat Q_1^{\text{ch}} = Q_1 \pm 2 e n_1\rangle$. 

In Appendix~\ref{app:secondTransformation}, we generalize the above approach to systems with more charge modes, where a more careful consideration of the flux periodicity in the circuit is needed to avoid double counting of states. In that case our basis for each charge mode is given by the eigenstates of that charge operator, and rescale the operators such that
\begin{equation}
    \label{eq:chargeQuantization}
    \hat{Q}^\text{ch}_i\ket{n_i} = (Q_i + 2en_i)\ket{n_i}.
\end{equation}
The gate charge $Q_i$ can affect the dynamics, and leads to changes in the energy levels, perhaps most dramatically in the charge limit of the Cooper pair box qubit.


We separate the basis of the transformed Hamiltonian into two subspaces. The first subspace is formed by $\nh$ harmonic oscillators and the second subspace is formed by $\nc$ charge modes, which is also the number of charge islands. Therefore, the general basis of the circuit is
\begin{equation}
    \label{eq:basis}
    \ket{n_1^{\text{ha}},\dots,n_{\nh}^{\text{ha}}}\ket{n_1^{\text{ch}},\dots,n_{\nc}^{\text{ch}}},
\end{equation}
where $n_m^{\text{ha}}=0,1,2,\cdots$ is number of photons in the $m$th harmonic mode and $n_m^{\text{ch}}=\cdots,-1,0,1,\cdots$ are the number of cooper pair that have tunneled onto the $m$th charge mode. 

We now express the Josephson potential energies in the basis chosen above. Using the fact that different flux operators act on different subspaces, the Josephson energies can be written as
\begin{equation*}
    \begin{split}
        &\cos \left(\frac{2\pi}{\Phi_0}\tilde{\bm{w}}^T_k\hat{\tilde{\bm{\Phi}}}+\bm{b}_k^T\bm{\varphi}_{\text{ext}}\right)= \\
        &\frac{1}{2}e^{i\bm{b}_k^T\bm{\varphi}_{\text{ext}}}\prod_{m=1}^{\nh}e^{i\frac{2\pi}{\Phi_0}w_{km}^{\text{ha}}\hat{\phi}^{\text{ha}}_m}\prod_{m=1}^{\nc}e^{i\frac{2\pi}{\Phi_0}w_{km}^{\text{ch}}\hat{\phi}^{\text{ch}}_m} + \text{h.c.},
    \end{split}
\end{equation*}
where $w_{km}^\lambda$ and $\hat{\phi}^\lambda_m$ are the $m$th element of $\bm{w}^\lambda_k$ and $\hat{\bm\Phi}^\lambda$ respectively for $\lambda \in \{\text{ha},\text{ch}\}$. By using annihilation and creation operator representation of $\hat{\phi}_k^{\text{ha}}$, Eq.~\ref{eq:fluxHarmonic}, we can write the exponential term related to harmonic modes as:
\begin{equation}
    e^{{i\frac{2\pi}{\Phi_0}w_{km}^{\text{ha}}\hat{\phi}^{\text{ha}}_m}}= \hat{D}_m(\alpha_{km}),
\end{equation}
where $\alpha_{km} = i\frac{2\pi }{\Phi_0}w_{km}^\text{ha}\sqrt{\frac{\hbar Z_m}{2}}$ and $\hat{D}_m(\alpha_{km})=\exp\left(\alpha_{km}a^\dagger_m - \alpha^*_{km}a_m\right)$ is bosonic displacement operator for harmonic mode $m$ whose off diagonal elements fall off quickly from the diagonal for small $|\alpha_{km}|\ll 1$. The exponential due to the charge mode is
\begin{equation}
    \begin{split}
        e^{i\frac{2\pi}{\Phi_0}w_{km}^{\text{ch}}\hat{\phi}^{\text{ch}}_m} = (\hat{d}_m)^{w_{km}^\text{ch}},
    \end{split}
\end{equation}
where $\hat{d}_m = \sum_{n_m^{\text{ch}}} |n_m^{\text{ch}} +1\rangle\langle n_m^{\text{ch}}|$ is the charge raising operator. Note that as a result of how we performed the circuit transformation, $w_{km}^\text{ch}$ is either $0$, $-1$ or $+1$. The operator raised to these powers corresponds to the identity, reducing the charge on island $m$ by one Cooper pair, or increasing it by a Cooper pair.

Finally, the transformed Hamiltonian of Eq.~\ref{eq:transfromedH} in the specified harmonic and charge operators has the following representation:
\begin{equation*}
    \small
    \begin{split}
        \hat{\tilde{H}} &= \hbar\sum_{m=1}^{\nh}\omega_m \hat{a}^\dagger_m \hat{a}_m + \frac{1}{2}\sum_{mn}(\bm{C}^{\text{ch}})^{-1}_{mn}\hat{Q}^{\text{ch}}_m\hat{Q}^{\text{ch}}_n\\
        &+\sum_{k\in \mathcal{S}_L}\left(\frac{\Phi_0}{2\pi}\frac{\bm{b}_k^T\bm{\varphi}_{\text{ext}}}{l_k}\right)\sum_{m=1}^{\nh} w_{km}^\text{ha}\sqrt{\frac{\hbar Z_m}{2}}(\hat{a}_m^\dagger + \hat{a}_m)\\
        &-\sum_{k\in\mathcal{S}_J}\frac{E_{J_k}}{2}\Big(e^{i\bm{b}_k^T\bm{\varphi}_{\text{ext}}}\prod_{m=1}^{\nh}\hat{D}_m(\alpha_{km})\prod_{m=1}^{\nc}(\hat{d}_m)^{w_{km}^\text{ch}}\\
        &+ \text{h.c.}\Big).
    \end{split}
\end{equation*}
To represent and diagonalize the Hamiltonian numerically, we must set the truncation number for each harmonic and charge mode. For example, if the truncation number of the second harmonic mode is $N$, the $n_2^\text{ha}$ in Eq.~\ref{eq:basis} can take any value from $0$ to $N-1$.

\begin{figure}[t]
  \centering
  \includegraphics[scale=0.45]{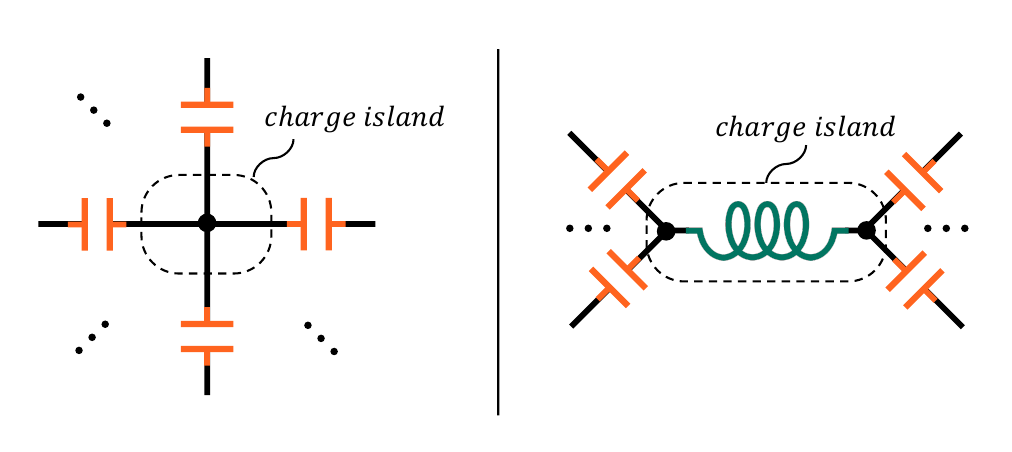}
  \caption{Examples of the conserved charge on the isolated island. Charges cannot escape the island because there is no Josephson junction connected to the island. \code{SQcicuit} detects these situations and removes them automatically.}
  \label{fig:charge}
\end{figure}

There are several key ideas associated to our method for the Hamiltonian diagonalization: (1) for the circuits only consist of linear elements such as capacitors and inductors, we show that proper coordinate transformation and basis selection is sufficient for diagonalizing the Hamiltonian, without a need to solve the eigenvalue problem; (2) for the circuits with nonlinearities, we employ the idea of finding and omitting the decouple modes  using the Josephson junction part of the Hamiltonian. This significantly reduces the dimension of the Hilbert space required for computing the spectrum. For example, these decouple modes frequently happen in symmetric circuits such as $0-\pi$ qubit (see Sec.~\ref{sec:SQcircuitIntro} for details). Another example of decoupled modes are conserved charges on isolated islands in superconducting circuits that cannot tunnel through Josephson junctions. These modes appear in the Hamiltonian with terms proportional to $\hat{Q}^2$ and do not have presence in the Josephson junction part of the Hamiltonian (see Fig.~\ref{fig:charge}). Since these charges are conserved, they have no dynamics, and we can remove them from the problem.

\section{Circuit Description in SQcircuit}\label{sec:SQcircuitIntro}
\begin{figure*}[t]
  \centering
  \includegraphics[scale=0.5]{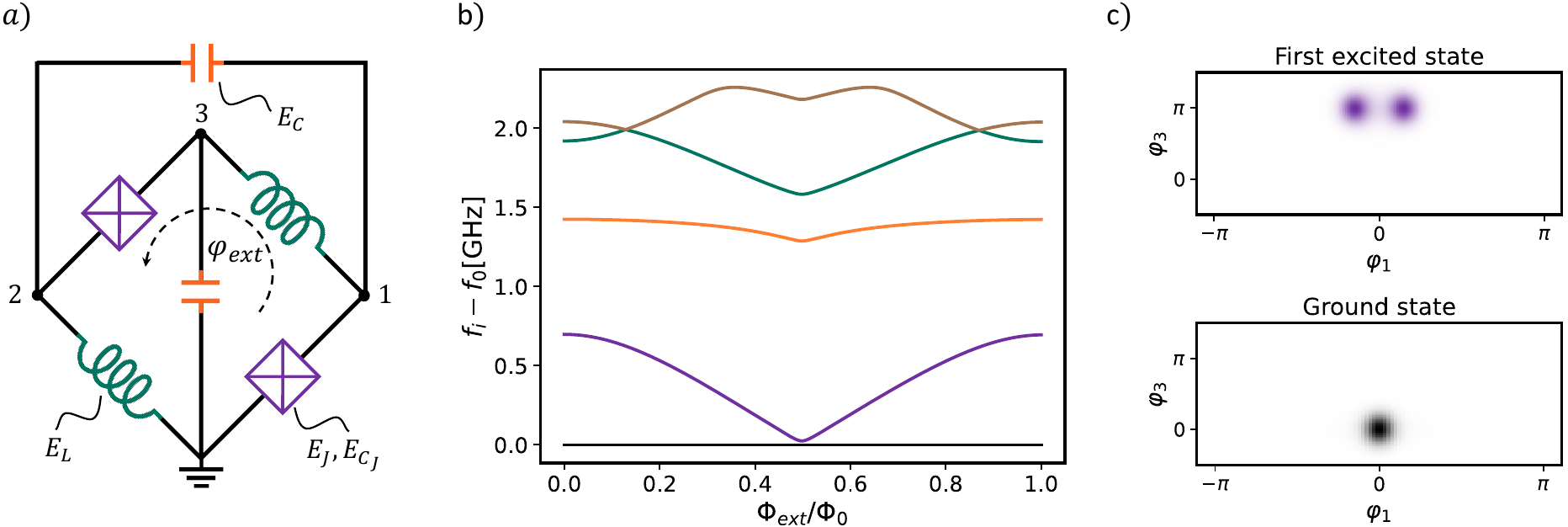}
  \caption{\textbf{a)} A symmetric $0-\pi$ qubit with inductive, Josephson, and charging energies in gigahertz unit as follow: $E_L/h=0.13~\text{GHz}$, $E_J/h=5~\text{GHz}$, $E_C/h=0.15~\text{GHz}$, and $E_{C_J}/h=10~\text{GHz}$. \textbf{b)} The energy spectrum of the circuit calculated by \code{diag()} functionality of the \code{SQcircuit}. \textbf{c)} The probability distribution of the ground and first excited states in phase space coordinates, calculated at $\varphi_\text{ext}=0$, by using \code{eig\_phase\_coord()} functionality of the \code{SQcircuit}. Figures are generated by the Python code in Appendix \ref{app:codes}.}
  \label{fig:zeroPi}
\end{figure*}
We have developed \code{SQcircuit}, a Python package that utilizes and automates the methods described in the previous sections to build and diagonalize the circuit Hamiltonian from its circuit description and extract experimentally relevant properties such as transition matrix elements and decay rates.  In this section, by using $0-\pi$ qubit~\cite{Groszkowski2017} shown in Fig.~\ref{fig:zeroPi}a as an example, we demonstrate how to describe a circuit in \code{SQcircuit} and to calculate its spectrum and eigenfunctions. We begin by importing the \code{SQcircuit} package:
\begin{lstlisting}[language=Python]
import SQcircuit as sq
\end{lstlisting}

In \code{SQcircuit}, each circuit component has an associated class definition and properties. 
To define the capacitors, we create an object of \code{Capacitor} class. For instance, the capacitors in Fig.~\ref{fig:zeroPi}a can be defined by
\begin{lstlisting}[language=Python]
C = sq.Capacitor(value=0.15, unit="GHz")
CJ = sq.Capacitor(value=10, unit="GHz") 
\end{lstlisting}
In the above lines of the code, the \code{unit} argument can be either hertz or farad. If the \code{unit} is in hertz, such as \str{"GHz"}, \str{"MHz"}, the \code{value} argument specifies the charging energy of the capacitor, i.e., $E_c = e^2/2c$. If the \code{unit} argument is in farad, such as \str{"nF"}, \str{"pF"}, etc., the value specifies the capacitance in farad. For example, the previous line of code for the \code{CJ} capacitor is approximately equivalent to
\begin{lstlisting}[language=Python]
CJ = sq.Capacitor(value=1.94, unit="fF")
\end{lstlisting}
since the $1.94$ femtofarad capacitor has a charging energy of approximately $10$ gigahertz.

Before defining the inductive elements, we need to define the inductive loops (closed path of inductive elements) within which the element resides and through which the flux will be set, and so we define the loops  for the circuit. For example, for the $0-\pi$ qubit, we define a single loop by creating an object of \code{Loop} class as
\begin{lstlisting}[language=Python]
loop1 = sq.Loop(value=0)
\end{lstlisting}
where \code{value} is the external flux value that can be altered later by \code{set\_flux()} method.

 The inductor in Fig.~\ref{fig:zeroPi}a can be created as an object from the \code{Inductor} class:
\begin{lstlisting}[language=Python]
L = sq.Inductor(value=0.13, unit="GHz",
                loops = [loop1])
\end{lstlisting}
Like the \code{Capacitor} class, if the \code{unit} argument is in hertz, the \code{value} describes the inductive energy, $E_l=(\Phi_0/2\pi)^2/l$. If the \code{unit} argument is in henry, such as \str{"uH"}, \str{"nH"}, etc., the \code{value} is the inductance in henry.  For inductive elements that are part of a loop, we indicate the loops in which they reside. For example, if the inductive element is part of the two loops, namely \code{loop1} and \code{loop2} (both objects of the Loop class), we would pass the arguments of \code{loops=[loop1, loop2]} to the definition of the inductive element. In our example, the inductor is only part of \code{loop1}. Thus, we set \code{loops=[loop1]}. The Josephson junction as an inductive element has an analogous definition to inductors and can be defined by using the \code{Junction} class. The Josephson junction in Fig.\ref{fig:zeroPi}a can be defined as
\begin{lstlisting}[language=Python]
JJ = sq.Junction(value=5, unit="GHz",
                 loops=[loop1])
\end{lstlisting}

The default unit for all elements is gigahertz.
However, we can change the default unit for each type of element via \code{sq.set\_unit\_cap()}, \code{sq.set\_unit\_ind()}, and \code{sq.set\_unit\_JJ()} functions for capacitors, inductors, and Josephson junctions respectively. For example the default unit of capacitors can be set to \str{"fF"} with
\begin{lstlisting}[language=Python]
sq.set_unit_cap("fF")
\end{lstlisting}

After defining all circuit components, to describe the circuit topology in \code{SQcircuit}, we create an object of \code{Circuit} class by passing a Python dictionary that contains the list of all elements at each edge. For the circuit of Fig.\ref{fig:zeroPi}a
\begin{lstlisting}[language=python]
elements = {
    (0, 1): [CJ, JJ],
    (0, 2): [L],
    (0, 3): [C],
    (1, 2): [C],
    (1, 3): [L],
    (2, 3): [CJ, JJ]
}
            
cr = sq.Circuit(elements)
\end{lstlisting}
By creating an object of \code{Circuit} class, \code{SQcircuit} systematically finds the set of transformations and bases for diagonalization of the Hamiltonian. Before setting the truncation numbers for each mode and diagonalizing the Hamiltonian, we can gain more insight into our circuit by calling the \code{description()} method. \code{SQcircuit} prints out the transformed Hamiltonian Eq.~\ref{eq:transfromedH} and a listing of the modes, whether they are harmonic or charge modes, and the frequency for each harmonic mode as given by  Eq.~\ref{eq:angularFreq}. Moreover, it shows the external flux distribution over inductive elements $\bm{b}^T_k$ and prefactors in the Josephson junction part of the Hamiltonian $\tilde{\bm{w}}^T_k$. For example, by executing it on the \code{cr} object:
\begin{lstlisting}[language=python]
cr.description()
\end{lstlisting}
\begin{equation*}\footnotesize
\begin{split}
&\hat{H} = \omega_1\hat{a}^\dagger_1\hat{a}_1 + E_{C_{22}}(\hat{n}_2-n_{g_2})^2-E_{J_1}\cos(\hat{\varphi}_1+\hat{\varphi}_2+\\
&0.5\varphi_{ext_1})-E_{J_2}\cos(\hat{\varphi}_1-\hat{\varphi}_2-0.5\varphi_{ext_1})\\
&-------------------------\\
&\text{mode 1:}~~~\text{harmonic}~~~\hat{\varphi}_1=\varphi_{zp_1}(\hat{a}_1+\hat{a}^\dagger_1)~~~\omega_1/2\pi= 3.22\\
&\text{mode 2:}~~~\text{charge}~~~~~~~n_{g_2}=0\\
&E_{C_{22}}=0.3~~E_{J_1}=5~~E_{J_2}=5~~~\varphi_{zp_1}=2.49\\
&-------------------------\\
\end{split}
\end{equation*}
The output above indicates that the first mode of the \code{cr} circuit is a harmonic mode with a natural frequency of $3.22$ gigahertz, as indicated by the frequency unit of \code{SQcircuit}. The second mode is a charge mode. There is a third mode that is not displayed in the Hamiltonian due to \code{SQcircuit} automatically eliminating decoupled modes. The normalized charge offset for the second mode ($n_{g_2}$) can be altered through \code{set\_charge\_offset()} method as
\begin{lstlisting}[language=python]
cr.set_charge_offset(mode=2, ng=1)
\end{lstlisting}
This method is useful for understanding the charge dispersion of the energy spectrum due to random offsets on charge islands~\cite{Koch2007}. 

The main step before diagonalizing the circuit is to set the truncation numbers. To do that, we pass a list of truncation numbers for each mode to the \code{set\_trunc\_nums()} method. For \code{cr} circuit:
\begin{lstlisting}[language=python]
cr.set_trunc_nums([25, 25])
\end{lstlisting}

Lastly, we use \code{diag()} to extract a specified number of eigenfrequencies and eigenvectors. We  find the first five eigenstates of the \code{cr} circuit via:
\begin{lstlisting}[language=python]
efreqs, evecs = cr.diag(n_eig=5)
\end{lstlisting}
where \code{n\_eig} specifies the number of eigenvalues to output. The lower \code{n\_eig}, the faster \code{SQcircuit} finds the eigenvalues. \code{efreqs} is a Numpy array that contains the eigenfrequencies in \code{SQcircuit} frequency units. \code{evecs} is a list that includes the eigenvectors, each as a QuTiP object. 

To generate the spectrum of the circuit, firstly, we need to change and sweep the external flux of \code{loop1} by the \code{set\_flux()} method. Then, we need to find the eigenfrequencies of the circuit that correspond to that external flux. The following lines of code find the flux tuning of the spectrum, \code{spec}, which is a 2D NumPy array where each column contains the eigenfrequencies with respect to the external flux. 
\begin{lstlisting}[language=python]
# array of external fluxes
phi = np.linspace(0, 1, 100)

# number of eigenvalues we aim for
n_eig = 5

# array that contains the spectrum
spec = np.zeros(n_eig, len(phi))

for i in range(len(phi)):

    # set the value of the flux
    loop1.set_flux(phi[i])
    
    # get the eigenfrequencies 
    spec[:, i], _ = cr.diag(n_eig)
\end{lstlisting}
The results are plotted in Fig.~\ref{fig:zeroPi}b (see Appendix \ref{app:codes}.) Note that the default unit of frequency in \code{SQcircuit} is gigahertz. We can change the default unit by calling \code{sq.set\_unit\_freq()} method. For example, we can change the unit to megahertz via
\begin{lstlisting}[language=python]
sq.set_unit_freq("MHz")
\end{lstlisting}

\section{Circuit Features and Properties}\label{sec:features}
We have developed a systematic way to construct a Hamiltonian for a superconducting quantum circuit and diagonalize it in a physically-motivated and efficient basis. However, for many quantum hardware design problems, the spectrum of the circuit is not sufficient. In this section, we extend the theory of  Sec.~\ref{sec:generalHamil} and \ref{sec:transCoord} to extract other essential circuit properties, and describe how to obtain those features using \code{SQcircuit}.

\subsection{Coordinate Representation of Wavefunctions}
\begin{figure*}[t]
  \centering
  \includegraphics[scale=0.5]{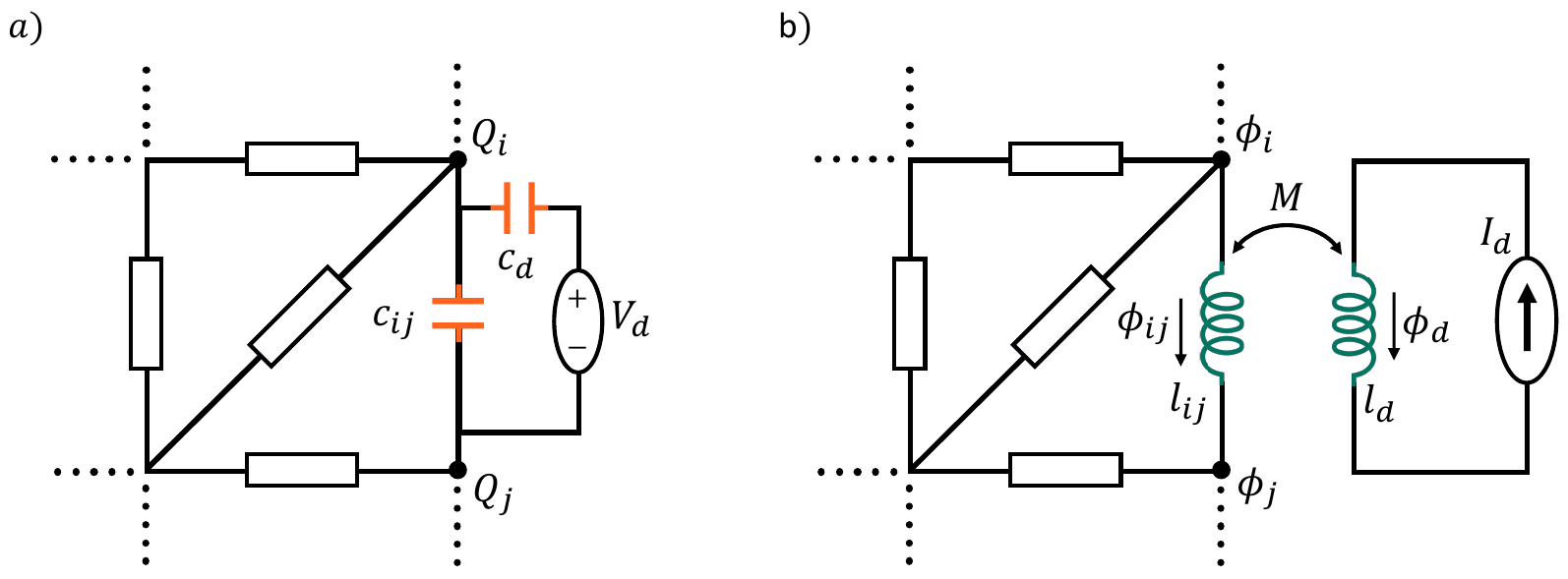}
  \caption{\textbf{a)} A drive voltage source coupled with small capacitance of $c_d$ to arbitrary nodes of $i$ and $j$ of a circuit. The $Q_i$ and $Q_j$ are the charge node operators of the node $i$ and $j$ respectively, and $C_{ij}$ is the capacitance between those two nodes. \textbf{b)} A current source coupled to an inductor between arbitrary nodes of $i$ and $j$ of a circuit with small mutual inductance of $M$. $\varphi_{ij}$ is the phase drop across the inductor $L_{ij}$ between nodes $i$ and $j$ with phase node operators of $\varphi_i$ and $\varphi_j$.}
  \label{fig:couplingCircuit}
\end{figure*}

Since many physical processes involve operators that are local in the flux (or equivalently phase $\varphi = 2\pi/\Phi_0\times\phi$), it is sometimes informative to plot the wavefunctions in this space. Given an eigenstate of the circuit in basis of Eq.~\ref{eq:basis}
\begin{equation*}
\small
    \begin{split}
        &\ket{\psi} =\\&\sum_{n_1^{\ha},\dots,n_{\nc}^{\ch}} C_{n_1^{\ha},\dots,n_{\nc}^{\ch}}    \ket{n_1^{\text{ha}},\dots,n_{\nh}^{\text{ha}}}\ket{n_1^{\text{ch}},\dots,n_{\nc}^{\text{ch}}},
    \end{split}
\end{equation*}
we are interested in finding the inner product in phase coordinates:
\begin{equation}
\label{eq:phaseSpaceEig}
\small
    \begin{split}
        &\braket{\varphi_1^{\ha}, \dots,\varphi_{\nh}^{\ha}, \varphi_1^{\ch}, \dots,\varphi_{N_c}^{\ch}|\psi} =\sum_{n_1^{\ha},\dots,n_{\nc}^{\ch}}C_{n_1^{\ha},\dots,n_{\nc}^{\ch}}\\
        &\braket{\varphi_1^{\ha}|n_1^{\ha}}\dots\braket{\varphi_{\nh}^{\ha}|n_{\nh}^{\ha}}\braket{\varphi_1^{\ch}|n_1^{\ch}}\dots\braket{\varphi_{\nc}^{\ch}|n_{\nc}^{\ch}}.
    \end{split}
\end{equation}
It suffices to express each harmonic and charge mode eigenstate in phase coordinate basis. In other words, we need to find $\braket{\varphi_m^\lambda|n_m^\lambda}$ for $\lambda\in\{\ha,\ch\}$. For the harmonic modes we use the phase coordinate representation of Fock states given by
\begin{equation}
\label{eq:phaseSpaceHarmonic}
\small
    \braket{\varphi^\text{ha}|n} = \frac{1}{\sqrt{\sqrt{\pi}2^n n!x_0}}e^{-(\frac{\Phi_0}{2\pi}\varphi^\text{ha})^2/2x_0^2}H_{n}\left(\frac{\Phi_0\varphi^\text{ha}}{2\pi x_0}\right),
\end{equation}
where $x_0 = \sqrt{\hbar Z_i}$ and $H_{n}$ are the Hermite polynomials. On the other hand, the phase coordinate representation of the charge mode eigenstates are given by 
\begin{equation}
\label{eq:phaseSpaceCharge}
\small
\braket{\varphi^{\ch}|n} = \frac{1}{\sqrt{2\pi}}\exp{(in\varphi^{\ch})}.
\end{equation}
The eigenvectors in the phase coordinate can be obtained by substituting the Eq.~\ref{eq:phaseSpaceHarmonic}-\ref{eq:phaseSpaceCharge} into Eq.~\ref{eq:phaseSpaceEig}. 

We implement the above basis transformation in \code{SQcircuit} within the function  \code{eig\_phase\_coord()}. For example, we obtain the ground state and first excited state of the $0-\pi$ qubit of Sec.~\ref{sec:SQcircuitIntro} at $\varphi_{ext}=0$ by executing the commands:
\begin{lstlisting}[language=python]
# create a range for each mode
phi1 = np.pi*np.linspace(-1,1,100)
phi2 = np.pi*np.linspace(-0.5,1.5,100)

# the ground state
state0 = cr.eig_phase_coord(
    k=0, 
    grid=[phi1, phi2]
)
    
# the first excited state
state1 = cr.eig_phase_coord(
    k=1, 
    grid=[phi1, phi2]
)
\end{lstlisting}
Here, \code{k} is an eigenvector index, which is $0$ for the ground state and $1$ for the first excited state. The argument \code{grid} is a list that specifies the values of phase $\varphi$ to evaluate the wavefunction at, which can be either a single number or a range of values for each mode. We therefore choose a range of values for \code{phi1} and \code{phi2}. 
The resulting phase coordinate representations of the states are \code{state0} and \code{state1}, which are $100\times100$-complex Numpy arrays holding the values of the eigenfunction evaluated on a grid corresponding to \code{phi1} and \code{phi2}. In Fig.~\ref{fig:zeroPi}c we plot $|\braket{\varphi_1,\varphi_2|\psi}|^2$ for the ground state \code{state0} and the first excited state \code{state1} (see Appendix \ref{app:codes} for more details). 

\subsection{Coupling Operators and Matrix Elements}

Having shown how to analyze an isolated circuit, we now move to describing our approach for understanding the interaction of a circuit with the outside world. More specifically, we aim to calculate the rates at which transitions occur between energy eigenstates under the effect of different driving mechanisms, such as the capacitive and inductive driving via control electronics. 
The same calculations also apply to estimate the effect of vacuum noise or other random fluctuations driving the system, enabling us to estimate quantities such as the qubit lifetime.

\paragraph{Capacitive coupling:} As shown in Fig.~\ref{fig:couplingCircuit}a, we can capacitively couple a voltage source to arbitrary nodes of $i$ and $j$ of a circuit. If the coupling capacitor, $c_d$, is small enough that the coupling elements do not change the spectrum of the primary circuit, the interaction Hamiltonian between the primary circuit and drive circuit will be (Appendix \ref{app:capDrive})
\begin{equation*}
    \hat{H}^\text{dr}_c = (c_dV_d) \bm{e}_{ij}^T\bm{C}^{-1}\bm{R}\hat{\tilde{\bm{Q}}},
\end{equation*}
where $V_d$ is a voltage of a source $\bm{e}_{ij}$ is the $\nn$ dimensional vector with $\delta_{ik} - \delta_{kj}$ for element $k\in\{1,2,\cdots,\nn\}$. We therefore define the capacitive coupling operator as 
\begin{equation}
    \label{eq:capacitiveCoupling}
    \hat{\mathcal{O}}_c \equiv \bm{e}_{ij}^T\bm{C}^{-1}\bm{R}\hat{\tilde{\bm{Q}}}.
\end{equation}
\paragraph{Inductive coupling:} The inductive coupling case, shown in Fig.~\ref{fig:couplingCircuit}b, is essentially a dual of the capacitive case and can be analyzed very similarly. Given an inductor $l_{ij}$ in the circuit, we introduce a mutual coupling $M$ which is small enough to have negligible effect on the main circuit's spectrum. The effect of the coupling is then captured by an interaction Hamiltonian (Appendix \ref{app:indDrive})
\begin{equation*}
    \hat{H}^\text{dr}_l = (MI_d) \frac{1}{l_{ij}}\bm{e}_{ij}^T\bm S\hat{\tilde{\bm\Phi}},
\end{equation*}
where $I_d$ is the source current. The inductive coupling operator is defined as 
\begin{equation}
    \label{eq:inductiveCoupling}
    \hat{\mathcal{O}}_l \equiv \frac{1}{l_{ij}}\bm{e}_{ij}^T\bm S\hat{\tilde{\bm\Phi}}.
\end{equation}

We now use the coupling operators defined above to estimate the transition rates between states. We first calculate the transition matrix element of each type of coupling operator between the energy eigenstates $\ket{m}$ and $\ket{n}$: 
\begin{equation}
    \label{eq:matrixElements}
    \bra{m} \hat{\mathcal{O}}_\lambda \ket{n},
\end{equation}
where $\lambda \in \{c, l\}$. In \code{SQcircuit}, the \code{coupling\_op()} method returns the coupling operator for each type of coupling. For example, to obtain $\hat{\mathcal{O}}_c$ as a QuTiP object in the constructed Fock/charge basis, we issue the command:
\begin{lstlisting}[language=python]
O_c = cr.coupling_op(ctype="capacitive",
                     nodes=(i, j))
\end{lstlisting}
The argument \code{ctype} sets the type of coupling to either \str{"capacitive"} or \str{"inductive"}, while the \code{nodes} argument is a Python tuple containing the nodes that couple to the drive circuit. Sometimes we are interested only in a specific element of the transition matrix element (Eq.~\ref{eq:matrixElements}) as opposed to the whole matrix. For this we use the \code{matrix\_elements()} method:
\begin{lstlisting}[language=python]
g_mn = cr.matrix_elements(ctype="capacitive",
                          nodes=(i, j),
                          states=(m, n))
\end{lstlisting}
where the states of interest are passed through the \code{states} argument.

\subsection{Estimating Decay and Dephasing Rates}
\begin{figure*}[t]
  \centering
  \includegraphics[scale=0.5]{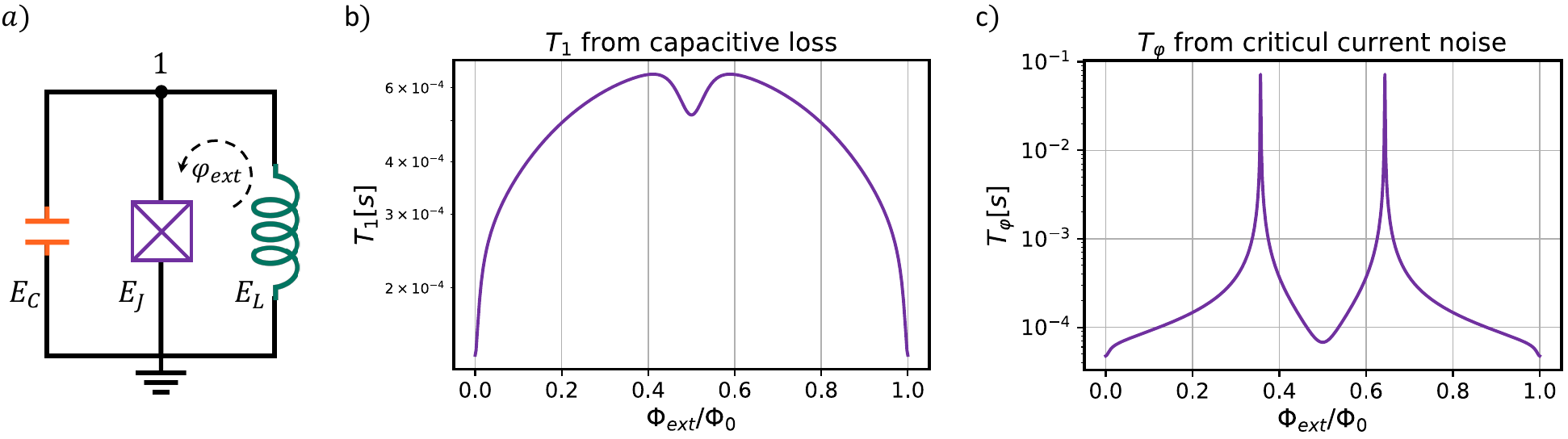}
  \caption{\textbf{a)} A Fluxonium qubit with inductive, Josephson, and charging energies in gigahertz unit as follow: $E_L/h=0.46~\text{GHz}$, $E_J/h=10.2~\text{GHz}$, and $E_C/h=3.6~\text{GHz}$. \textbf{b},\textbf{c)} $T_1$ and $T_\varphi$ due to capacitive loss and critical current noise of a Fluxonium with $Q_{cap} = 10^6$ and $A_{cc}/E_J = 5\times10^{-7}$ is plotted as a function of external flux by using \code{dec\_rate()} functionality of the \code{SQcircuit}. Appendix \ref{app:codes} contains the Python code that produces these figures.}  
  \label{fig:loss}
\end{figure*}
We need to understand the effects of imperfect components and fluctuating parameters on the resulting dynamics in the quantum description of superconducting circuits. In this section, we show how \code{SQcircuit} can take properties -- such as the dissipation present in the materials within capacitors, losses in inductors, the presence of quasiparticles in junctions,  parasitic coupling to output channels, and  fluctuations in parameters -- and convert them into dissipation rates and jump operators that enter a master equation description of the open quantum system's dynamics.

\subsubsection{Depolarization}
Consider a circuit that is coupled to a degree of freedom of the environment $\hat{\xi}$ with the interaction Hamiltonian
\begin{equation*}
    H_{\text{int}} =\hat{\mathcal{O}}\hat{\xi}.
\end{equation*}
The operator $\hat{\mathcal{O}}$ is in the circuit Hilbert space, while $\hat{\xi}$ is an operator acting on the environment. Within the context of linear response theory~\cite{Clerk2010},  we find the decay rate
\begin{equation}
    \label{eq:rateDecoherence}
    \Gamma_{m \rightarrow n} = \frac{1}{\hbar^2}|\bra{m}\hat{\mathcal{O}}\ket{n}|^2S_{\xi\xi}(\omega_{mn}),
\end{equation}
from level $m$ to $n$. The transition frequency $\omega_{mn}= \omega_m - \omega_n$ is positive when the circuit emits energy to the environment and negative when circuit absorbs energy from it. The spectral density function (SDF) $S_{\xi\xi}(\omega_{mn})$ is defined as
\begin{equation*}
    {S_{\xi\xi}(\omega)} = \int_{-\infty}^\infty d\tau\braket{\xi(\tau)\xi(0)}e^{-i\omega\tau},
\end{equation*}
and satisfies the detailed balance relation~\cite{Clerk2010} of 
\begin{equation*}
    \frac{S_{\xi\xi}(\omega)}{S_{\xi\xi}(-\omega)} = \exp{\left(\frac{\hbar\omega}{k_BT}\right)},
\end{equation*}
where $k_B$ is the Boltzmann’s constant and $T$ is the bath temperature. In \code{SQcircuit}, the default bath temperature is $0.015$ kelvin, and it can be changed it by 
\begin{lstlisting}[language=python]
cr.set_temp(T=0.010)
\end{lstlisting}
where \code{cr} is an object of the \code{Circuit} class. 

To make quantitative predictions, we need to specify both operator $\hat{\mathcal{O}}$ and the SDF $S_{\xi\xi}(\omega)$. \code{SQcircuit} includes three common loss channels below that are currently considered to be significant in state-of-the-art circuits.

\paragraph{Capacitive Loss:} 
Dissipation in the capacitor dielectric can be modeled by a capacitor connected to a bath with voltage operator $\hat{V}$. For the capacitance $c_{ij}$ between nodes of $i$ and $j$, the circuit operator is
\begin{equation*}
    \hat{\mathcal{O}} = c_{ij}\bm{e}_{ij}^T\bm{C}^{-1}\bm{R}\hat{\tilde{\bm{Q}}},
\end{equation*}
with spectral density function of
\begin{equation*}
S_{VV}(\omega) = \frac{\hbar}{c_{ij}Q_{\text{cap}}(\omega)}\left(1+\coth{\frac{\hbar \omega}{2k_BT}}\right),   
\end{equation*}
where $Q_{\text{cap}}(\omega)$ is a quality factor which may be weakly frequency dependent. 
To calculate $\Gamma_{m \rightarrow n}$, the SDF $S_{VV}(\omega)$ is evaluated at the relevant transition frequency $\omega_{mn}$.
In \code{SQcircuit} the default capacitor quality factor is~\cite{Braginsky1987}
\begin{equation}
    \label{eq:Q_cap}
    Q_{\text{cap}}(\omega) =10^6\left(\frac{2\pi\times6\text{GHz}}{|\omega|}\right)^{0.7}, 
\end{equation}
but this can be set independently for each capacitor in the circuit, either as a scalar value or a function of $\omega$. For example, to set $Q_{\text{cap}}(\omega)=2\times10^6$ for the capacitor of the fluxonium in Fig.\ref{fig:loss}a, we define the capacitor with
\begin{lstlisting}[language=python]
C = sq.Capacitor(3.6, "GHz", Q=2e6)
\end{lstlisting}

\paragraph{Inductive Loss:} This is a loss due to quasiparticle tunneling across Josephson junctions of the superinductors and highly depends on their design. To model this loss, we can assume that the inductor between $i$ and $j$ nodes is coupled to a bath with current $\hat{I}$. The circuit operator $\hat{\mathcal{O}}$, hence, is equivalent to 
\begin{equation*}
    \hat{\mathcal{O}} = \bm{e}_{ij}^T\bm S\hat{\tilde{\bm\Phi}},
\end{equation*}
and
\begin{equation*}
    S_{II}(\omega) = \frac{\hbar}{l_{ij}Q_{\text{ind}}(\omega)}\left(1+\coth{\frac{\hbar \omega}{2k_BT}}\right),
\end{equation*}
where $Q_{\text{ind}}(\omega)$ is a frequency-dependent quality factor, with the following default value in \code{SQcircuit}~\cite{Smith2020}:
\begin{equation*}
    Q_{\text{ind}}(\omega) =500\times10^6\frac{K_0(\frac{h\times0.5\text{GHz}}{2k_BT})\sinh{(\frac{h\times0.5\text{GHz}}{2k_BT})}}{K_0(\frac{\hbar|\omega|}{2k_BT})\sinh{(\frac{\hbar|\omega|}{2k_BT})}},
\end{equation*}
in which $K_0$ is the Bessel function of the second kind. 
More generally, $Q_{\text{ind}}(\omega)$ can be set to any desired value or function, and
for the Fluxonium inductor in Fig.\ref{fig:loss}a with $Q_{\text{ind}} = 300\times10^6$, we could define
\begin{lstlisting}[language=python]
L = sq.Inductor(0.46, "nH", Q=300e6)
\end{lstlisting}

\paragraph{Quasiparticle Loss:} This loss occurs due to quasiparticle tunnelling in each Josephson junction. Based on ~\cite{Catelani2011}, the circuit operator for Josephson junction at the edge $k$ is 
\begin{equation}
    \hat{\mathcal{O}} = \sin{\left(\frac{1}{2}\left(\frac{2\pi}{\Phi_0}\tilde{\bm{w}}^T_k\hat{\tilde{\bm{\Phi}}}-\bm{b}_k^T\bm{\varphi}_{\text{ext}}\right)\right)}\footnote{To calculate the matrix element of $\hat{O}$ in the charge basis, we need to transform the Cooper pair charge basis to the single electron charge basis, similar to the approach in \cite{Aumann2021}.},
\end{equation}
and SDF is
\begin{equation*}
    S_{\qp}(\omega) = \hbar\omega \text{Re}[Y_{\qp}(\omega)]\left(1+\coth{\frac{\hbar \omega}{2k_BT}}\right),
\end{equation*}
where real part of admittance is 
\begin{multline*}
    \text{Re}[Y_{\qp}(\omega)]= \sqrt{\frac{2}{\pi}}\frac{8E_{J_k}}{R_\text{K}\Delta}\left(\frac{2\Delta}{\hbar\omega}\right)^{3/2}x_{\qp}\sqrt\frac{\hbar|\omega|}{2k_BT}\\ \times K_0\left(\frac{\hbar|\omega|}{2k_BT}\right)\sinh{\left(\frac{\hbar|\omega|}{2k_BT}\right)},
\end{multline*}
in which $R_\text{K}=h/e^2$ is the von Klitzing constant, $\Delta$ is the superconducting gap, and $x_{\qp}$ is the quasiparticle density. \code{SQcircuit} has default value of $\Delta=3.4\times10^{-4}~\text{eV}$ and $x_{\qp}=3\times10^{-6}$, but these values can be altered when defining the circuit Josephson junctions. For the Fluxonium in Fig.\ref{fig:loss}a, we change $\Delta$ and $x_{\qp}$ to different values via
\begin{lstlisting}[language=python]
JJ = sq.Junction(10.2, "GHz",
                 delta=2.5e-4, x=8e-6)
\end{lstlisting}

Finally, after defining all lossy elements and diagonalizing the circuit, the decay rate of Eq.~\ref{eq:rateDecoherence} can be calculated with
\begin{lstlisting}[language=python]
Gamma_mn = cr.dec_rate(dec_type=dec_type, 
                       states=(m, n),
                       total=total)
\end{lstlisting}
In the above code, \code{dec\_type} argument specifies the decay types and can be either \str{"capacitive"}, \str{"inductive"}, or \str{"quasiparticle"}. If \code{total=False}, \code{SQcircuit} calculates only the downward decay rate $\Gamma_{m\rightarrow n}$ if $m>n$, or $\Gamma_{n\rightarrow m}$ if $n>m$. If \code{total=True} which is the default value, \code{SQcircuit} returns the summation of both downward and upward transitions, that is $\frac{1}{T_1}= \Gamma_{m\rightarrow n}+\Gamma_{n\rightarrow m}$. As an example, Fig.\ref{fig:loss}b plots $T_1$ of the Fluxonium qubit due to capacitive loss as a function of external fluxes. The Python code that generates this plot is provided in Appendix \ref{app:codes}.

\subsubsection{Dephasing}
Besides causing transitions between different eigenstates, noise processes also lead to dephasing errors.
More specifically, if the transition frequency $\omega_{mn}(\lambda)$ depends on some external parameter $\lambda = \lambda_0 + \delta\lambda(t)$ where $\lambda_0$ is the desired value and $\delta\lambda(t)$ is some noise process with $\braket{\delta\lambda(t)}=0$, then information about
the relative phase between the $m$th and $n$th eigenstates is degraded due to fluctuations in $\lambda$.
Superconducting circuits are mainly affected by $1/f$ noise such as flux, charge, and critical current noise. The SDF for the $1/f$ noises is given by 
\begin{equation*}
    S_{\lambda\lambda}(\omega) = \frac{2\pi A_\lambda^2}{|\omega|},
\end{equation*}
where $A_\lambda$ is the noise amplitude. The pure dephasing time and its rate between $m$th and $n$th eigenstates is given by ~\cite{Groszkowski2017, Ithier2005}
\begin{equation}
    \label{eq:dephasingRate}
    \small
    \begin{split}
        \frac{1}{T_\phi}=\kappa_{mn}=\Big(2A_\lambda^2(\frac{\partial \omega_{mn}}{\partial\lambda})^2|\ln{\omega_\text{low}t_\text{exp}}|\\
        +2A^4_\lambda(\frac{\partial^2 \omega_{mn}}{\partial\lambda^2})^2(\ln^2{\frac{\omega_\text{hi}}{\omega_\text{low}}}+2\ln^2{\omega_\text{low}t_\text{exp}})\Big)^{1/2},
    \end{split}
\end{equation}
where $\omega_\text{low}$ and $\omega_\text{hi}$ are the low-frequency and high-frequency cutoff and $t_\text{exp}$ is the measurement time. The default value for those parameters in \code{SQcircuit} are as follow:
$\omega_\text{low}=1\text{Hz}$, $\omega_\text{hi}=3.0\text{GHz}$, and $t_\text{exp}=10\mu s$ which can be changed by
\begin{lstlisting}[language=python]
cr.set_low_freq(value, unit)
cr.set_high_freq(value, unit)
cr.set_t_exp(value, unit)
\end{lstlisting}
where \code{cr} is an object of \code{Circuit} class. Below we give details on how \code{SQcircuit} treats the three major $1/f$ noise channels that contribute to dephasing.

\paragraph{Critical current noise:} This noise is due to fluctuations in the Josephson energy. In \code{SQcircuit}, the default value for the normalized noise amplitude is $A \equiv A_J/E_J = 10^{-7}$, and a different value such as $A=5\times10^{-7}$ can be specified in the junction definition (Fig.\ref{fig:loss}a) with
\begin{lstlisting}[language=python]
JJ = sq.Junction(10.2, "GHz", x=8e-6, 
                 delta=2.5e-4, A=5e-7)
\end{lstlisting}
\paragraph{Charge noise:} This noise is due to randomness of the charge in the circuit charge islands. \code{SQcircuit} has the default normalized noise amplitude of $A \equiv A_{n_g}/e = 10^{-4}$, which can be modified for each charge mode of the circuit via \code{set\_charge\_noise()} method
\begin{lstlisting}[language=python]
cr.set_charge_noise(mode=1, A=2e-4)
\end{lstlisting}
Here \code{mode} argument specifies the index of the charge mode. To see which circuit mode is a charge mode, one can use \code{description()} method explained in Sec.~\ref{sec:SQcircuitIntro}.
\paragraph{Flux noise:} This noise is due to the fluctuations of the external fluxes coupled to inductive loops of the circuit. $A\equiv A_{\varphi_\text{ext}}/2\pi=10^{-6}$ is the default value for the normalized noise amplitude and can be altered in inductive loop definition with 
\begin{lstlisting}[language=python]
loop1 = sq.Loop(value=0, A=5e-6)
\end{lstlisting}
As discussed in ~\cite{You2019, Riwar2021} and Appendix \ref{app:timeDependentFlux}, it is vital to correctly specify the distribution of the external fluxes over inductive elements to properly calculate the dephasing rate for the circuit. This requires knowing the capacitor associated with each inductive elements, which can be specified via the \code{cap} argument in the inductor and junction definitions in \code{SQcircuit}. It also requires to pass \code{flux\_dist=}\str{"all"} option to \code{Circuit()} class definition. For the Fluxonium in Fig.\ref{fig:loss}a, we could define the circuit with following codes:
\begin{lstlisting}[language=python]
# user-defined flux noise 
loop1 = sq.Loop(value=0, A=5e-6)

C = sq.Capacitor(3.6, "GHz", Q=1e6)
L = sq.Inductor(0.46, "GHz", loops=[loop1])

# assign C to JJ
JJ = sq.Junction(10.2, "GHz", cap=C,
                 A=5e-7, loops=[loop1])
                 
# C is inside the JJ
elements = {(0, 1): [L, JJ]}

# define the Fluxonium
cr = sq.Circuit(elements, flux_dist="all")
\end{lstlisting}
where we assumed that the inductor has a tiny capacitor and the Josephson junction has \code{C} capacitor. Notice that since we have assigned \code{C} to \code{JJ}, there is no need to add \code{C} to the edge list of \code{elements}.

Finally, to calculate the dephasing rate of Eq.~\ref{eq:dephasingRate}, we again use the \code{dec\_rate()} method:
\begin{lstlisting}[language=python]
kappa_mn = cr.dec_rate(dec_type=dec_type, 
                       states=(m, n))
\end{lstlisting}
where \code{dec\_type} argument can be either, \str{"cc"}, \str{"flux"}, or \str{"charge"}. In Fig.\ref{fig:loss}c, we plot the dephasing time $T_\phi$ due to the critical current noise of the Fluxonium qubit with the Python code in Appendix \ref{app:codes}.

\section{Online presence}
We published \code{SQcircuit} as an open source Python package under BSD-3 license, which discloses that it is free to use and to be developed. The main source code can be found under the ~\cite{sqGit} GitHub repository, which provides the link to download and install the software. In addition, we prepared numerous examples that reproduce the result of the state of the art superconducting circuits in the literature to show the efficiency and convenience of the \code{SQcircuit} usage. We also provided an online documentation over ~\cite{sqWeb} for detailed information and tutorial of the \code{SQcircuit}. The Sphinx code that generate this documentation is made available on different GitHub repository of ~\cite{sqDocGit}. The list of the online links related to GitHub pages, website, documentation, and Anaconda and PyPi package
index repository pages are summarized at table ~\ref{tab:links}. We welcome any feedback regarding bugs/comments/improvements to the \code{SQcircuit} software packages as well as any initial pull request on the corresponding GitHub pages. If you find this useful, you are encouraged to cite this paper, where the framework has been developed.

\begin{table}[h!]
\footnotesize
\centering
{\rowcolors{0}{backcolour}{}
\bgroup
\def\arraystretch{1}
\begin{tabular}{ m{7.8cm} }
\pbox{7.7cm}{SQcircuit website: \\ 
\link{sqcircuit.org}}
\rule[-1.3em]{0pt}{2.8em}\\
\pbox{7.7cm}{SQcircuit online documentation: \\
\link{docs.sqcircuit.org}}
\rule[-1.3em]{0pt}{2.8em}\\
\pbox{7.7cm}{GitHub repository for SQcircuit source code:\\
\link{github.com/stanfordLINQS/SQcircuit}}
\rule[-1.3em]{0pt}{2.8em}\\
\pbox{7.7cm}{GitHub repository for Sphinx source code:\\
\link{github.com/stanfordLINQS/SQcircuit-doc}}
\rule[-1.3em]{0pt}{2.8em}\\
\pbox{7.7cm}{GitHub repository for Jupyter notebook examples:\\
\link{github.com/stanfordLINQS/SQcircuit-examples}}
\rule[-1.3em]{0pt}{2.8em}\\
\pbox{7.7cm}{Anaconda package repository page:\\
\link{anaconda.org/conda-forge/sqcircuit}}
\rule[-1.3em]{0pt}{2.8em}\\
\pbox{7.7cm}{PyPi package repository page:\\
\link{pypi.org/project/SQcircuit/}}
\rule[-1.3em]{0pt}{2.8em}\\

\end{tabular}}
\egroup
\caption{Online links related to website, GitHub pages, as well as package
index repository pages of \code{SQcircuit}.}
\label{tab:links}
\end{table}

\section{Conclusions and Future Works}\label{sec:concolusion}
In this work, we propose a general approach for analyzing any arbitrary superconducting circuit. We theoretically develop a systematic framework that models the Hamiltonian of the circuit. Then, we show a proper transformation of the coordinates that efficiently diagonalize the Hamiltonian to model its quantum behavior. We also extend our frameworks to capture other important features of the circuits such as lifetime, matrix elements, coupling operators, phase space representation of eigenfunctions, etc. Numerically, we introduce the \code{SQcircuit} library: an open-source Python library enabling the simulation and analysis of any superconducting circuit. We explain its functionalities to diagonalize the circuit Hamiltonian and to obtain the circuit features such as lifetime, matrix elements. Future work will aim to broaden the scope of the \code{SQcircuit} by adding more loss channels, creating a framework for optimization and machine learning approaches to discover and design sophisticated superconducting circuits, and including new functionalities to build a composite circuit created out of established superconducting circuits.

\section*{Note added}

During preparation of this manuscript, we were made aware of ongoing work in Jens Koch's group with a similar scope. Our two groups did not discuss details of our projects, but agreed to post preprints on the arXiv simultaneously.

\section*{Acknowledgments}

A.H.S.-N. is indebted to Dr. Marek Pechal for invaluable discussions. 
We acknowledge funding by Amazon Web Services Inc. and by the U.S. government through the Office of Naval Research (ONR) under grant No. N00014-20-1-2422 and the National Science Foundation CAREER award No.~ECCS-1941826. The authors wish to thank NTT Research for their financial and technical support and also acknowledge support by Keysight Inc. through the SystemX program at Stanford.
\newpage
\appendix
\section{Example for Circuit Hamiltonian}\label{app:exampleCircuit}
\begin{figure*}[t]
  \centering
  \includegraphics[scale=0.4]{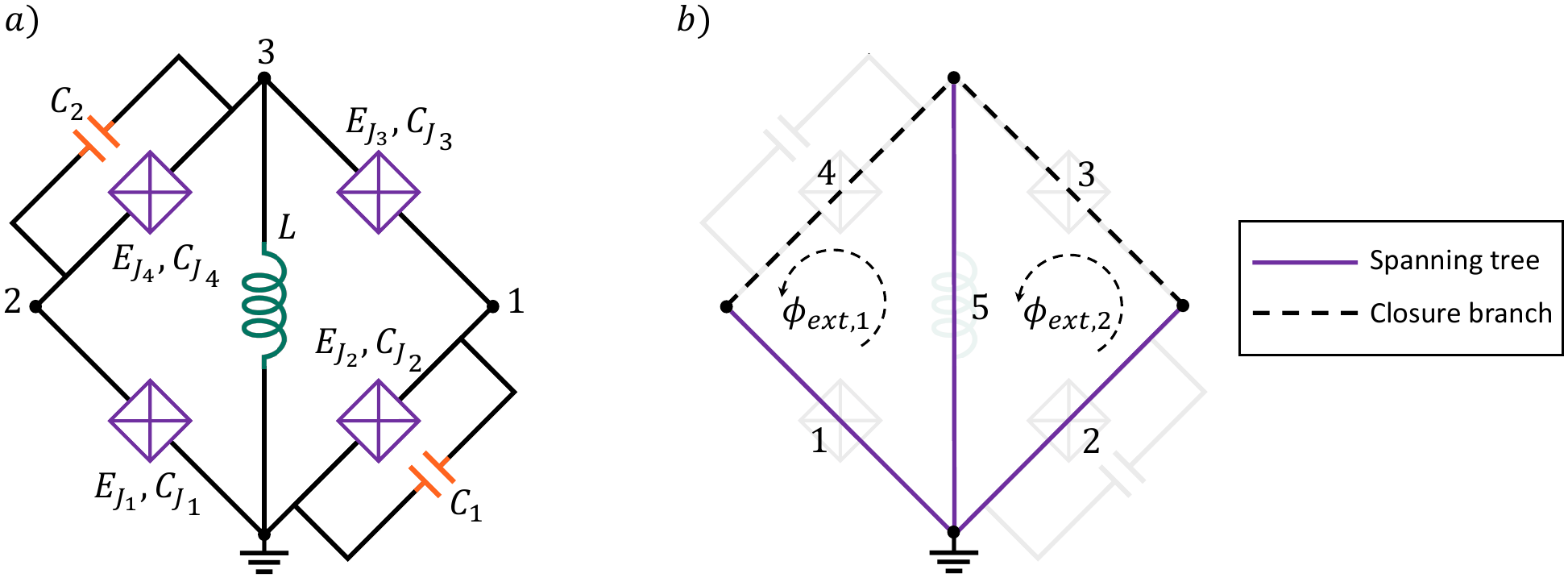}
  \caption{\textbf{a)} An arbitrary superconducting circuit with two inductive loops. \textbf{b)} The spanning tree of the circuit is specified by purple solid lines and closure branches are the black dashed lines. Branch numbers are specified next to each line.}
  \label{fig:example}
\end{figure*}

To elaborate upon the theory discussed in Sec.~\ref{sec:generalHamil}, we construct the Hamiltonian for the circuit shown in Fig.~\ref{fig:example}. To obtain the Hamiltonian of the form shown in Eq.~\ref{eq:circuitH}, we specify $\bm{C}$, $\bm{L}^*$, $\bm{w}_k$, and $\bm{b}_k$.  To construct the capacitance matrix, we set the diagonals elements to be the sum of the capacitance connected to a node and the off-diagonal elements to be negative the sum of capacitance between two nodes:  
\begin{equation*}
\small
    \bm{C} = 
    \begin{bmatrix}
     c_1+c_{J_2}+c_{J_3}  &  0 & - c_{J_3} \\
     0 &  c_2+c_{J_1}+c_{J_4} & -c_2-c_{J_4}  \\
     - c_{J_3} &  -c_2-c_{J_4} & c_2+c_{J_3} 
    \end{bmatrix}.\\
\end{equation*}
We construct the inverse inductance or susceptance matrix $\bm{L}^*$ similarly, by setting diagonal elements to be the sum of the susceptance connected to a node and off-diagonal elements to be negative the sum of susceptance between two nodes: 
\begin{equation*}
    \bm{L}^{*} = 
    \begin{bmatrix}
     0  &  0 & 0 \\
     0 &  0 & 0  \\
     0 & 0 & \frac{1}{l}
    \end{bmatrix}.\\
\end{equation*}
To calculate $\bm{w}_k$ and $\bm{b}_k$ defined in Eq.~\ref{eq:branchFlux}, we find a relation between the $k$th inductive branch flux, node fluxes, and external fluxes by specifying a minimum spanning tree. In Fig. \ref{fig:example}b the purple solid lines specify the chosen spanning tree and black dashed lines specify closure branches. The index of each branch is also specified; thus, the Josephson junction index set is $\mathcal{S}_J=\{1,2,3,4\}$ and linear inductor index set is $\mathcal{S}_L=\{5\}$. The Josephson junction on branch $k=1$ is part of the spanning tree, so its branch flux includes no offset from external fluxes:
\begin{equation*}
    \phi_{b,1} = (\phi_1 - 0) + 0\times\phi_{\ext,1} + 0\times\phi_{\ext,2}.
\end{equation*}
Hence, 
\begin{equation*}
    \bm{w}_1^T = \begin{bmatrix}1  &  0 & 0 \end{bmatrix},~~~~\bm{b}_1^T = \begin{bmatrix}0  &  0\end{bmatrix}.
\end{equation*}
In contrast, the Josephson junction on branch $k=3$ is not part of the spanning tree, but is on a closure branch that encloses $\phi_{\ext,2}$. Therefore the branch flux depends on $\phi_{\ext,2}$ as, 
\begin{equation*}
    \phi_{b,1} = (\phi_1 - \phi_3) + 0\times\phi_{\ext,1} + 1\times\phi_{\ext,2},
\end{equation*}
which specifies $\bm{w}_3$ and $\bm{b}_3$:
\begin{equation*}
    \bm{w}_3^T = \begin{bmatrix}1  &  0 & -1 \end{bmatrix},~~~~\bm{b}_3^T = \begin{bmatrix}0  &  1 \end{bmatrix}.
\end{equation*}
Using the same approach, the $\bm{w}_k$ and $\bm{b}_k$ for other branches are
\begin{equation*}
    \begin{split}
        \bm{w}_2^T = \begin{bmatrix}0  &  1 & 0 \end{bmatrix},~~~~\bm{b}_2^T = \begin{bmatrix}0  &  0  \end{bmatrix},\\
        \bm{w}_4^T = \begin{bmatrix}0  &  1 & -1 \end{bmatrix},~~~~\bm{b}_4^T = \begin{bmatrix}1  &  0 \end{bmatrix},\\
        \bm{w}_5^T = \begin{bmatrix}0  &  0 & 1 \end{bmatrix},~~~~\bm{b}_5^T = \begin{bmatrix}0  &  0 \end{bmatrix}.
    \end{split} 
\end{equation*}
We note that this approach for specifying $\bm{b}_k$ via spanning tree is complete only for time-independent external fluxes ~\cite{You2019, Riwar2021}. A generalized description including time-dependent external fluxes is provided in Appendix \ref{app:timeDependentFlux}.

\section{Coordinate Transformations}\label{app:coordinateTransformation}
In this section we systematically construct transformed  capacitance and inverse inductance matrices as described in Sec.~\ref{sec:transCoord}, to obtain the form in Eqs.~\ref{eq:transformedC} and \ref{eq:transformedL}. The transformation occurs in two steps. In the first step we diagonalize the $LC$ part of the Hamiltonian and separate the Hilbert space into two subspaces of harmonic and charge modes. In the second step we transform only the charge-mode subspace to simplify the boundary conditions and representation of the charge modes. Thus, we write the $\bm{S}$ and $\bm{R}$ matrices as,
\begin{align*}
    \bm{R} =  \bm{R}_1\bm{R}_2,\\
    \bm{S} =  \bm{S}_1\bm{S}_2,
\end{align*}
where $\bm{R}_\lambda$ and $\bm{S}_\lambda$ specify the $\lambda$th transformation.
\subsection{First Transformation}
We write the first transformation as:
\begin{align}
    \label{eq:transformedC1}
    \bm{C}^{-1}_1 = \bm{R}_1^T\bm{C}^{-1} \bm{R}_1,\\
    \bm{L}^{*}_1 = \bm{S}_1^T\bm{L}^{*}\bm{S}_1.
\end{align}
The goal is to find the  $\bm{R}_1$ and $\bm{S}_1$, such that $\bm{C}_1$ and $\bm{L}^{*}_1$ are diagonal. By taking the inverse of Eq.~\ref{eq:transformedC1} and imposing the transformation is canonical, \ie, $\bm{R}_1^T = \bm{S}_1^{-1}$, we obtain:
\begin{align}
    \label{eq:transformedC1L1}
    \begin{split}
        \bm{C}_1 = \bm{S}_1^T\bm{C} \bm{S}_1,\\
        \bm{L}^{*}_1 = \bm{S}_1^T\bm{L}^{*}\bm{S}_1.
    \end{split}
\end{align}
Therefore, the problem reduces to finding the matrix $\bm{S}_1$ such that $\bm{C}_1$ and $\bm{L}^{*}_1$ are diagonal. Because ($\bm{C}$, $\bm{L}^{*}$) are symmetric and positive-semidefinite, we can take their square root and rewrite $\bm{C}$ and $\bm{L}^{*}$ as~\cite{Pechal2017}
\begin{equation*}
    \begin{split}
        \bm{C} = \sqrt{\bm{C}}^T\sqrt{\bm{C}}, \\
        \bm{L}^{*} = \sqrt{\bm{L}^{*}}^T\sqrt{\bm{L}^{*}}.
    \end{split}
\end{equation*}
By substituting in Eq.~\ref{eq:transformedC1L1} we find
\begin{equation}\label{eq:C1L1asFcFl}
    \begin{split}
        \bm{C}_1 = \bm{S}_1^T\sqrt{\bm{C}}^T\sqrt{\bm{C}}\bm{S}_1 = \bm{F}_c^T\bm{F}_c,\\
        \bm{L}^{*}_1 = \bm{S}_1^T\sqrt{\bm{L}^{*}}^T\sqrt{\bm{L}^{*}}\bm{S}_1= \bm{F}_l^T\bm{F}_l,
    \end{split}
\end{equation}
where 
\begin{equation}\label{eq:F1}
    \begin{split}
        \bm{F}_c = \sqrt{\bm{C}}\bm{S}_1,\\
        \bm{F}_l = \sqrt{\bm{L}^{*}}\bm{S}_1.
    \end{split}
\end{equation}
Because $\bm{C}_1$ and $\bm{L}^{*}_1$ are assumed to be diagonal, Eq.~\ref{eq:C1L1asFcFl} implies that
\begin{equation}\label{eq:F2}
    \begin{split}
        \bm{F}_c = \bm{U}\bm{D}_c,\\
        \bm{F}_l = \bm{V}\bm{D}_l,
    \end{split}
\end{equation}
where $\bm{D}_c$ and $\bm{D}_l$ are diagonal matrices and $\bm{U}$ and $\bm{V}$ are orthogonal matrices. By combining Eq.~\ref{eq:F1} and Eq.~\ref{eq:F2}, we have:
\begin{align}
    \label{eq:C_S1_UD}
    \sqrt{\bm{C}}\bm{S}_1 = \bm{U}\bm{D}_c,\\
    \label{eq:L_S1_UD}
    \sqrt{\bm{L}^{*}}\bm{S}_1 = \bm{V}\bm{D}_l.
\end{align}
Because $\bm{C}$ is invertible, we can find $\bm{S}_1$ by multiplying the left side of Eq.~\ref{eq:C_S1_UD} by $\sqrt{\bm{C}}^{-1}$:
\begin{equation}\label{eq:S1}
    \bm{S}_1 = \sqrt{\bm{C}}^{-1}\bm{U}\bm{D}_c
\end{equation}
Substituting Eq.~\ref{eq:S1} in Eq.~\ref{eq:L_S1_UD}:
\begin{equation}\label{eq:svd}
    \sqrt{\bm{L}^{*}}\sqrt{\bm{C}}^{-1} = \bm{V}\bm{D}_l\bm{D}_c^{-1}\bm{U}^T= \bm{V}\bm{D}\bm{U}^T,
\end{equation}
where 
\begin{equation}\label{eq:D}
    \bm{D}=\bm{D}_l\bm{D}_c^{-1}.
\end{equation}
The right side of Eq.~\ref{eq:svd} has the form of a singular value decomposition (SVD). Hence, we can specify $\bm{S}_1$ from $\bm{U}$ and $\bm{D}_c$ matrices. We are free to choose  $\bm{D}_c$ and $\bm{D}_l$ as long as they satisfy Eq.~\ref{eq:D}. However, the zero diagonal entries of $\bm{D}$ imply that the corresponding diagonal entries of $\bm{D}_l$ should also be zero because $\bm{D}_c$ is invertible and cannot have zero diagonal entries. In addition, permuting the $\bm{D}$ diagonal entries can create another acceptable form of SVD. We sort the $\bm{D}$ diagonal entries to place the zero entries at the last indices, defining a $2 \times 2$ block matrix partition in which we call the first $\nh$ non-zero diagonal elements the ``harmonic'' block and the last $\nc$ zero elements the ``charge'' block. Thus, $\bm{C}_1$ and $\bm{L}^{*}_1$ have the following form:
\begin{equation*}
    \begin{split}
        \bm{C}_1 = \bm{D}_c^2 = \begin{bmatrix}
    \bm{C}^{\ha_1}   &   \bm{0}\\
    \bm{0}   &   \bm{C}^{\ch_1}
    \end{bmatrix},\\
        \bm{L}^{*}_1 = \bm{D}_l^2=\begin{bmatrix}
    \bm{L}^{*^{\ha_1}}   &   \bm{0}\\
    \bm{0}   &   \bm{0}
    \end{bmatrix},
    \end{split}
\end{equation*}
where $\bm{C}{^\ha_1}$ and $\bm{L}^{*^\ha_1}$ are $\nh\times \nh$ diagonal matrices and $\bm{C}^{\ch_1}$ is a $\nc\times \nc$ diagonal matrix.

\subsection{Second Transformation}\label{app:secondTransformation}

Since the elements of $\bm{L}^{*}_1$  are zero for the charge block, there is no quadratic potential for this subspace. Instead, the Josephson junction cosine potentials form a $\nc$-dimensional periodic potential:
\begin{equation}\label{eq:periodicPotential}
    U(\bm{\Phi}^{\ha_1},\bm{\Phi}^{\ch_1}+\bm{a}) = U(\bm{\Phi}^{\ha_1},\bm{\Phi}^{\ch_1}),
\end{equation}
where $\bm{\phi}^{\ha_1^T}=[\phi^{\ha_1}_1, \dots, \phi^{\ha_1}_{\nh}]$ and $\bm{\Phi}^{\ch_1^T}=[\phi^{\ch_1}_1, \dots, \phi^{\ch_1}_{\nc}]$ are flux variables after the first transformation for the harmonic and charge subspaces respectively, and $\bm{a}$ is an $\nc$-dimensional vector defined as 
\begin{equation*}
    \bm{a} = \sum_{i=1}^{\nc} m_i\bm{a}_i,~~ m_i\in\mathbb{Z},
\end{equation*}
where $\bm{a}_i$ are primitive lattice vectors. We can write the periodic potential of Eq.~\ref{eq:periodicPotential} as a Fourier series:
\begin{equation} \label{eq:fourierPotential}
    U(\bm{\Phi}^{\ha_1},\bm{\Phi}^{\ch_1}) = \sum_{\bm{k}} c_{\bm{k}}(\bm{\Phi}^{\ha_1})e^{\frac{i}{\hbar}\bm{k}^T\bm{\Phi}^{\ch_1}},
\end{equation}
where $\bm{k}$ is a reciprocal lattice vector formed from basis vectors $\bm{k}_i$ satisfying $\bm{k}_i^T\bm{a}_j=2\pi\hbar\delta_{ij}$. 
\medskip

Consider $\ket{\bm{Q}^{\ch_1}_0}=\ket{Q^{\ch_1}_1, \dots, Q^{\ch_1}_{\nc}}$, an eigenvector of all charge operators $\hat{Q}^{\ch_1}_i$ in the charge block after the first transformation, with
\begin{equation}\label{eq:chargeBlockEigVec}
    \hat{Q}^{\ch_1}_i\ket{\bm{Q}^{\ch_1}_0} = Q^{\ch_1}_i\ket{\bm{Q}^{\ch_1}_0}.
\end{equation}
Because $[\hat{\phi}^{\ch_1}_m ,\hat{Q}^{\ch_1}_n]=i\hbar\delta_{mn}$, the operator $e^{\frac{i}{\hbar}\bm{k}^T\bm{\hat{\Phi}}^{\ch_1}}$ performs a charge displacement on $\ket{\bm{Q}^{\ch_1}_0}$:
\begin{equation} \label{eq:chargeDispOp}
    e^{\frac{i}{\hbar}\bm{k}^T\hat{\bm{\Phi}}^{\ch_1}}\ket{\bm{Q}^{\ch_1}_0} = \ket{\bm{Q}^{\ch_1}_0+\bm{k}}.
\end{equation}
As a consequence of Eqs.~\ref{eq:fourierPotential}-\ref{eq:chargeDispOp}, the Hilbert space corresponding to the charge block is spanned by a discrete set of charge eigenvectors $\ket{\bm{Q}^{\ch_1}}$ on a reciprocal lattice. The lattice so far is defined up to a reference vector which we can take to be $\ket{\bm{Q}^{\ch_1}_0}$, defining the following basis for the charge-block subspace:
\begin{equation*}
\mathcal{H}_{\bm{Q}^{\ch_1}_0} = \{\ket{\bm{Q}^{\ch_1}_0+\bm{k}}|\bm{k}=\sum_{i=1}^{\nc}m_i\bm{k_i}, m_i\in\mathbb{Z}\}.
\end{equation*}
Any vector $\bm{k}$ on the reciprocal lattice can be absorbed into the definition of $\ket{\bm{Q}^{\ch_1}_0}$, so we can take $\ket{\bm{Q}^{\ch_1}_0}$ to describe a possible non-integer offset to the reciprocal lattice. The offset, if present, corresponds to a conserved quantity that is equivalent to a quasi-momentum and is often called "gate charge" or "charge offset".

For calculations, charge operators and the potential of Eq.~\ref{eq:fourierPotential} have a more physically meaningful representations if each reciprocal vector has the the following form: 
\begin{equation}
    \label{eq:transformedReciprocal}
    \tilde{\bm{k}}_i = (2e)\bm{e}_i,
\end{equation}
where $2e$ is the Cooper pair charge and $\bm{e}_i$ is the $\nc$-dimensional vector with the $i$th element equal to one and the rest equal to zero (standard basis). We therefore construct $\bm{S}_2$ and $\bm{R}_2$ for the second transformation such that Eq.~\ref{eq:transformedReciprocal} is satisfied. The transformation acts only on the charge block, so $\bm{S}_2$ and $\bm{R}_2$ are block-diagonal:
\begin{align*}
    \bm{S}_2 = \begin{bmatrix}
    \bm{I}   &   \bm{0}\\
    \bm{0}   &   \bm{S}^{\ch}_2
    \end{bmatrix},\\
    \bm{R}_2 = \begin{bmatrix}
    \bm{I}   &   \bm{0}\\
    \bm{0}   &   \bm{R}^{\ch}_2
    \end{bmatrix},\\
\end{align*}
where $\bm{I}$ is an $\nh\times \nh$ identity matrix and $(\bm{S}^{\ch}_2)^T = (\bm{R}^{\ch}_2)^{-1}$. Under these transformations Eq.~\ref{eq:fourierPotential} becomes,
\begin{equation*}
        U(\bm{\Phi}^{\ha_2},\bm{\Phi}^{\ch_2}) = \sum_{\bm{k}} c_{\bm{k}}(\bm{\Phi}^{\ha_2})e^{\frac{i}{\hbar}\bm{k}^T\bm{S}^{\ch}_2\bm{\Phi}^{\ch_2}},
\end{equation*}
where $\bm{\Phi}^{\ha_2}$ and $\bm{\Phi}^{\ch_2}$ are the flux variables after the second transformation, and 
$\bm{\Phi}^{\ha_2} = \bm{\Phi}^{\ha_1}$. Hence, the new reciprocal vectors $\tilde{\bm{k}}_i$ are related to previous ones by
\begin{equation*}
    \tilde{\bm{k}}^T_i = \bm{k}^T_i\bm{S}^{\ch}_2.
\end{equation*}
To ensure the $\tilde{\bm{k}}^T_i $ have the form of Eq.~\ref{eq:transformedReciprocal}, $\bm{S}^{\ch}_2$ should satisfy:
\begin{equation*}
    \bm{S}^{\ch}_2 = (\bm{K}^{T})^{-1},
\end{equation*}
 where the $i$th column of the $\bm{K}$ matrix is $\bm{k}_i$. 
 \medskip
 
 So far we have considered only the generic form of a periodic potential in Eq.~\ref{eq:periodicPotential}. In practice, the reciprocal lattice vectors $\bm{k}_i$ for the Josephson potential are proportional to the prefactor vectors $\bm{w}_i^{\ch_1}$ within the cosines after the \textit{first} transformation, defined by truncating $\bm{S}_1^{T}\bm{w}_i$ to the charge block. We can therefore form the columns of $\bm{K}$ using the $\nc$ linearly-independent $\bm{w}_i^{\ch_1}$:
 \begin{equation}\label{eq:WtoK}
    \bm{K} = (2e) \begin{bmatrix} \bm{w}_1^{\ch_1} & \dots & \bm{w}_{\nc}^{\ch_1} \end{bmatrix}
 \end{equation}
 This guarantees that all $\bm{w}_k^{\ch_2}$ (prefactors of the flux variables after the \textit{second} transformation) are integers, and that the reciprocal vectors have the form of Eq.~\ref{eq:transformedReciprocal}. Thus, after the second transformation the capacitance and inductance matrices  $\bm{C}_2$ and $\bm{L}^{*}_2$ are
\begin{equation*}
    \begin{split}
        \bm{C}_2 = \begin{bmatrix}
    \bm{C}^{\ha_2}   &   \bm{0}\\
    \bm{0}   &   \bm{C}^{\ch_2}
    \end{bmatrix},\\
        \bm{L}^{*}_2 =\begin{bmatrix}
    \bm{L}^{*^\ha_2}   &   \bm{0}\\
    \bm{0}   &   \bm{0}
    \end{bmatrix},
    \end{split}
\end{equation*}

 where $\bm{C}^{\ha_1} = \bm{C}^{\ha_2}$, $\bm{L}^{*^\ha_2} = \bm{L}^{*^\ha_1}$, and $\bm{C}^{\ch_2} = (\bm{S}^{\ch}_2)^T\bm{C}^{\ch_1}\bm{S}^{\ch}_2$. Although $\bm{C}^{\ch_1}$ is diagonal, $\bm{C}^{\ch_2}$ is not necessarily diagonal, so $\bm{C}_2$ and $\bm{L}^{*}_2$ have the same form as Eqs.~\ref{eq:transformedC}-\ref{eq:transformedL}.
 
\section{Interaction Hamiltonian}
Here we obtain interaction Hamiltonians for the two generic circuits shown in Fig. \ref{fig:couplingCircuit}. To do so, we describe how the drive circuit changes the Lagrangian of the main circuit, then compute the Hamiltonian by Legendre transform. We derive additional terms describing capacitive drives and inductive drives in Sec.~\ref{app:capDrive} and Sec.~\ref{app:indDrive}, respectively.

\subsection{Capacitive Drive}
\label{app:capDrive}
The coupling capacitor and voltage drive of Fig. \ref{fig:couplingCircuit}a only affect the kinetic energy of the circuit as follows,
\begin{align}
    \label{eq:newKeneticCap}
    T^\prime(\dot{\bm{\Phi}}) = T(\dot{\bm{\Phi}}) + \frac{1}{2}c_d(\dot\phi_i - \dot\phi_j - V_d)^2,\\
    \label{eq:newPotentialCap}
    U^\prime(\bm{\Phi}) = U(\bm{\Phi}),
\end{align}
where $T(\dot{\bm{\Phi}})=\frac{1}{2}\bm{\dot\Phi}^T \bm{C} \bm{\dot\Phi}$ and $U(\bm{\Phi})$ are respectively the kinetic and potential energies related to the circuit without drive components, and ($\phi_i$, $\phi_j$) are flux variables for the nodes $i$ and $j$ that are connected through the drive circuit. The Lagrangian is $\mathcal{L} = T - U$, from which we compute redefined charge operators (as momenta conjugate to the flux variables):
\begin{equation}
    \label{eq:newCharge}
    \small
    Q^{\prime}_t = \frac{\partial}{\partial \dot\phi_t}\mathcal{L} = \sum_{m}C_{tm}\dot\phi_m + 
    c_d(\delta_{it}-\delta_{jt})(\dot\phi_i - \dot\phi_j - V_d).
\end{equation} 
We define the new capacitance matrix $\bm{C}^\prime$ and charge drive vector $\bm{q_d}$ with following elements
\begin{equation*}
    \begin{split}
        C^\prime_{mn} = C_{mn} + c_d(\delta_{im}-\delta_{jm})\delta_{mn},\\
        q_{d_m} = c_d V_d (\delta_{im}-\delta_{jm}),
    \end{split}
\end{equation*}
and rewrite  Eqns .~\ref{eq:newKeneticCap} and ~\ref{eq:newCharge} in those parameters:
\begin{align}
    \label{eq:newCharge_2}
    \bm{Q}^\prime = \bm{C}^\prime\dot{\bm{\Phi}} - \bm{q_d},\\
    \label{eq:newKeneticCap_2}
    T^\prime= \frac{1}{2}\dot{\bm{\Phi}}^T C^\prime \dot{\bm{\Phi}} - \bm{q_d}^T\dot{\bm{\Phi}} + \frac{1}{4}\bm{q_d}^T\frac{1}{c_d}\bm{q_d}.
\end{align}
The Hamiltonian of the circuit is
\begin{equation}
    \label{eq:hamilCap}
    H = \bm{Q}^{\prime^T}\bm{\dot\Phi} -  T^\prime(\dot{\bm{\Phi}}) + U^\prime(\bm{\Phi}),
\end{equation}
Thus, by substituting Eqns.~\ref{eq:newPotentialCap}, \ref{eq:newCharge_2}, \ref{eq:newKeneticCap_2} in Eq.~\ref{eq:hamilCap}:
\begin{equation*}
    \begin{split}
       H =& \frac{1}{2}\bm{Q}^{\prime^T}\bm{C}^{\prime^{-1}}\bm{Q}^\prime+ U(\bm{\Phi}) +\bm{q_d}^T\bm{C}^{\prime^{-1}}\bm{Q}^\prime \\
       &+ \frac{1}{2}\bm{q_d}^T\bm{C}^{\prime^{-1}}\bm{q_d} - \frac{1}{4}\bm{q_d}^T\frac{1}{c_d}\bm{q_d}.
    \end{split}
\end{equation*}
The last two terms contain no dynamical variables, so we neglect them as they describe only a global shift to the energy spectrum.  Assuming $c_d$ is much smaller than capacitors connected to nodes $i$ and $j$, we can substitute the $\bm{Q}^\prime$ and $\bm{C}^\prime$ with $\bm{Q}$ and $\bm{C}$. Accordingly, the Hamiltonian can be written as
\begin{equation}\label{eq:finalDrive2}
    \hat{H} = \hat{H}_0 + \bm{q_d}^T\bm{C}^{-1}\bm{\hat{Q}},
\end{equation}
where $\hat{H}_0$ is the Hamiltonian of the circuit without drives. The interaction Hamiltonian for capacitive driving can be written using transformed charge operators as:
\begin{equation*}
    \hat{H}^\text{dr}_c = (c_dV_d) \bm{e}_{ij}^T\bm{C}^{-1}\bm{R}\hat{\tilde{\bm{Q}}},
\end{equation*}
where we used $\bm{q}_d=c_dV_d\bm{e}_{ij}$ and $\bm{e}_{ij}$ is the $\nn$- dimensional vector with element $k\in\{1,2,\cdots,\nn\}$ equal to $\delta_{ik} - \delta_{kj}$.

\subsection{Inductive Drive}
\label{app:indDrive}
In contrast to capacitive coupling, inductive coupling only changes the potential of the circuit and leaves the kinetic energy intact. The potential energy of the inductor $l_{ij}$ coupled to the drive circuit in Fig.\ref{fig:couplingCircuit}b is,
\begin{equation}
    \label{eq:inductiveEnergy}
    \Delta U = \frac{1}{2}l_{ij} I_{ij}^2 + MI_dI_{ij} + \frac{1}{2}l_{d} I_d^2,
\end{equation}
where $I_{ij}$ is the current through the $l_{ij}$ inductor and $I_d$ is the source current. The flux drop across the $l_{ij}$ and $l_{d}$ inductors has the following relation with their currents due to mutual inductance $M$:
\begin{equation}
    \label{eq:phaseDrop}
    \begin{split}
         \phi_{ij} = l_{ij}I_{ij} + MI_d,\\
         \phi_d = MI_{ij} + l_d I_d,
    \end{split}
\end{equation}
Assuming $M\ll l_{ij}$, we express the Eq.~\ref{eq:inductiveEnergy} in $\phi_{ij}$ using Eq.~\ref{eq:phaseDrop}:
\begin{equation*}
    \Delta U = \frac{1}{2l_{ij}}\phi_{ij}^2 + \frac{MI_d}{l_{ij}}\phi_{ij} + \frac{1}{2}l_{d} I_d^2.
\end{equation*}
Thus, the kinetic and the potential energy of the circuit are
\begin{equation}\label{eq:kenAndPot}
    \begin{split}
        T^\prime(\dot{\bm{\Phi}}) = T(\dot{\bm{\Phi}}),\\
        U^\prime(\bm{\Phi}) = U(\bm{\Phi})+ \frac{MI_d}{l_{ij}}\phi_{ij} + \frac{1}{2}l_{d} I_d^2,
    \end{split}
\end{equation}
in which  $T(\dot{\bm{\Phi}})$ and $U(\bm{\Phi})$ are the kinetic and potential energy respectively related to the circuit without drive components. By neglecting $\frac{1}{2}l_{d}I_d^2$  (since it is just a shift to energy spectrum of a circuit), the Hamiltonian can be obtained from the Lagrangian formed by Eq.~\ref{eq:kenAndPot} as 
\begin{equation*}
    \hat{H} = \hat{H}_0 + \frac{MI_d}{l_{ij}}\bm{e}_{ij}^T\hat{\bm{\Phi}},
\end{equation*}
where $\hat{H}_0$ is the Hamiltonian of the circuit without drive elements, and we used $\hat{\phi}_{i} - \hat{\phi}_{j}=\bm{e}_{ij}^T\hat{\bm{\Phi}}$. Note that if the $i$ or $j$ is the ground node (that is $i=0$) the corresponding flux operator is zero ($\hat{\phi}_{i}$ = 0). The interaction Hamiltonian in transformed flux operators from inductive drive is:
\begin{equation*}
    \hat{H}^\text{dr}_l = (MI_d) \frac{1}{l_{ij}}\bm{e}_{ij}^T\bm S\hat{\tilde{\bm\Phi}}.
\end{equation*}

\section{Time-dependent External Fluxes}\label{app:timeDependentFlux}
Different forms of spanning tree lead to different distributions $\bm{b}_k$ of the external fluxes over the inductive elements, and therefore to different forms of the Hamiltonian. As discussed in ~\cite{You2019, Riwar2021}, for stationary external fluxes, all distributions $\bm{b}_k$ lead to Hamiltonians with the same energy spectrum. However, for time-dependent external fluxes terms proportional to $\propto\dot{\phi}\dot{\varphi}_\ext$ appear in the kinetic energy part of the Lagrangian; this is not captured in the description of Sec.~\ref{sec:generalHamil}. In this section, we explain a method to find the $\bm{b}_k$ for a Hamiltonian with time-dependent external fluxes, that remains consistent with the time-independent case in Sec.~\ref{sec:generalHamil}.  Let all branch fluxes be denoted by 
\begin{equation*}
    \bm{\Phi}^T_b=[\phi_{b,1},\dots,\phi_{b,(\nn+\nl)}]
\end{equation*}
where $\phi_{b,k}$ is the $k$th branch flux, $\nn$ is the number of circuit nodes, and $\nl$ is the number of independent loops ($\nl+\nn$ is the number of branches). Fluxoid quantization and Faraday’s law determine constraints for the $k$th loop $\mathcal{A}_k$ threaded by external flux $\phi_{\ext,k}$:
\begin{equation}\label{eq:loopCondition_k}
    \phi_{\ext,k} = \bm{g}_k^T\bm{\Phi}_b,
\end{equation}
where 
\begin{equation*}
\footnotesize
 [\bm{g}_k]_i =
 \begin{cases}
    \text{ 1},\text { if }\phi_{b,i} \in \mathcal{A}_k\text{ with same direction as }\phi_{\ext,k}&\\
    \text{-1},\text{ if }\phi_{b,i} \in \mathcal{A}_k\text{ with opposite direction to }\phi_{\ext,k}&\\
    \text{ 0},\text{ if }\phi_{b,i} \notin \mathcal{A}_k&
 \end{cases}
\end{equation*}
Thus, Eq.~\ref{eq:loopCondition_k} for all $\nl$ loops of the circuit can be written as 
\begin{equation}\label{eq:loopCondition}
    \bm{\Phi}_\ext = \bm{G}\bm{\Phi}_b,
\end{equation}
where $\bm{G}$ is a $\nl\times (\nn+\nl)$ matrix with $k$th row equal to $\bm{g}_k^T$. Moreover, the relation between $k$th branch flux $\phi_{b,k}$, node flux variables $\bm{\Phi}$ (circuit degrees of freedom), and external fluxes $\bm{\Phi}_\ext$ can be expressed by, 
\begin{equation}\label{eq:edgePhi}
    \phi_{b,k} = \bm{w}_k^T\bm{\Phi} + \bm{b}_k^T\bm{\Phi}_\ext,
\end{equation}
where $\bm{w}_k$ are calculated as in Sec.~\ref{sec:generalHamil} and $\bm{b}_k$ are yet to be specified. Eq.~\ref{eq:edgePhi} for all $k$ can be written in the matrix form of
\begin{equation}\label{eq:edgePhiVec}
    \bm{\Phi}_b = \bm{W}\bm{\Phi} + \bm{B}\bm{\Phi}_\ext,
\end{equation}
where $\bm{W}$ and $\bm{B}$ are $(\nn+\nl)\times \nn$  and $(\nn+\nl)\times \nl$  matrices with respective $k$th rows given by $\bm{w}_k^T$ and $\bm{b}_k^T$. We define an $(\nn+\nl)$-dimensional vector $\bm{\Phi}^{+}$ and $(\nn+\nl)\times(\nn+\nl)$ matrix of $\bm{P}$ as, 
\begin{equation*}
    \begin{split}
        \bm{\Phi}^{+} =  \begin{bmatrix}\bm{\Phi}  \\ \bm{\Phi}_\ext\end{bmatrix},\\
        \bm{P} =\begin{bmatrix}\bm{W}  & \bm{B}\end{bmatrix},
    \end{split}
\end{equation*}
to rewrite Eq.~\ref{eq:edgePhiVec} compactly as
\begin{equation}\label{eq:compactPhiEdge}
    \bm{\Phi}_b = \bm{P}\bm{\Phi}^{+}.
\end{equation}
The kinetic energy of the Lagrangian is,
\begin{equation*}
    T = \frac{1}{2}\dot{\bm{\Phi}}_b^T\bm{C}_\text{ed}\dot{\bm{\Phi}}_b,
\end{equation*}
where $\bm{C}_\text{ed}$ is an $(\nn+\nl)\times(\nn+\nl)$ diagonal matrix with $k$th diagonal element equal to the capacitance of the $k$th branch. Using Eq.~\ref{eq:compactPhiEdge}, the kinetic energy can be written as
\begin{equation*}
    T = \frac{1}{2}(\dot{\bm{\Phi}}^{+})^T\bm{C}\dot{\bm{\Phi}}^{+},
\end{equation*}
where,
\begin{equation*}
    \begin{split}
    \bm{C} &= \bm{P}^T\bm{C}_\text{ed}\bm{P}=
    \begin{bmatrix}\bm{W}^T  \\ \bm{B}^T\end{bmatrix}\bm{C}_\text{ed}\begin{bmatrix}\bm{W}  & \bm{B}\end{bmatrix}
    \\
    &=\begin{bmatrix}\bm{W}^T\bm{C}_\text{ed}\bm{W}& \bm{W}^T\bm{C}_\text{ed}\bm{B}\\
    \bm{B}^T\bm{C}_\text{ed}\bm{W} & \bm{B}^T\bm{C}_\text{ed}\bm{B}
    \end{bmatrix}
    \end{split}
\end{equation*}
The off-diagonal blocks of $\bm{C}$ lead to terms $\propto\dot{\phi}\dot{\varphi}_\ext$. To avoid this, the $\bm{B}$ matrix should satisfy the following condition:
\begin{equation}\label{eq:conditionOnB_1}
    \bm{W}^T\bm{C}_\text{ed}\bm{B} = \bm{0},
\end{equation}
where $\bm{0}$ is a $\nn\times \nl$ zero matrix. The $\bm{B}$ matrix has $(\nn+\nl)\times \nl$ elements that are unknown and Eq.~\ref{eq:conditionOnB_1} introduces $\nn\times \nl$ number of equations. Thus, $\nl^2$ equations are needed to specify $\bm{B}$ uniquely. Substituting Eq.~\ref{eq:compactPhiEdge} in Eq.~\ref{eq:loopCondition}, we have:
\begin{equation*}
    \begin{split}
        \bm{G}\bm{\Phi}_b &= \bm{G}\bm{P}\bm{\Phi}^{+} =\bm{G}\begin{bmatrix}\bm{W}  & \bm{B}\end{bmatrix} \begin{bmatrix}\bm{\Phi}  \\ \bm{\Phi}_\ext\end{bmatrix}\\
        &=\bm{G}\bm{W}\bm{\Phi}+\bm{G}\bm{B}\bm{\Phi}_\ext.
    \end{split}
\end{equation*}
$\bm{G}\bm{W}$ should be zero by definition, since by summing the flux drop at each branch of the loop, the net result should be independent of the flux node operators. Hence,
\begin{equation*}
    \bm{G}\bm{B}\bm{\Phi}_\ext = \bm{\Phi}_\ext.
\end{equation*}
Because this equation should hold for all $\bm{\Phi}_\ext$, we find the following constraint for $\bm{B}$, which provides the $\nl^2$ needed equations:
\begin{equation}\label{eq:conditionOnB_2}
    \bm{G}\bm{B}= \bm{I},
\end{equation}
where $\bm{I}$ is an $\nl\times \nl$ identity matrix. We combine Eqs.~\ref{eq:conditionOnB_1} and ~\ref{eq:conditionOnB_2} as 
\begin{equation*}
    \begin{bmatrix}\bm{W}^T\bm{C}_\text{ed}  \\ \bm{G}\end{bmatrix}\bm{B} = 
    \begin{bmatrix}\bm{0}  \\ \bm{I}\end{bmatrix},
\end{equation*}
and specify $\bm{B}$ by 
\begin{equation*}
    \bm{B} = 
    \begin{bmatrix}\bm{W}^T\bm{C}_\text{ed}  \\ \bm{G}\end{bmatrix}^{-1}\begin{bmatrix}\bm{0}  \\ \bm{I}\end{bmatrix}.
\end{equation*}
This process has been implemented in \code{SQcircuit} to describe time-dependent external fluxes, in particular to accurately predict dephasing rates due to flux noise.

\section{SQcircuit Codes}\label{app:codes}
Here we provide Python code that uses the \code{SQcircuit} library to produce results in the main body of this paper. 
We obtain the results of Fig.\ref{fig:zeroPi} by 
\begin{lstlisting}[language=python]
# import the libraries
import SQcircuit as sq
import numpy as np
import matplotlib.pyplot as plt

# define the loop of the circuit
loop1 = sq.Loop()

# define the circuit's elements
C = sq.Capacitor(0.15, "GHz")
CJ = sq.Capacitor(10, "GHz")
JJ = sq.Junction(5, "GHz", loops=[loop1])
L = sq.Inductor(0.13, "GHz", loops=[loop1])

# define the zero-pi circuit
elements = {
    (0, 1): [CJ, JJ],
    (0, 2): [L],
    (0, 3): [C],
    (1, 2): [C],
    (1, 3): [L],
    (2, 3): [CJ, JJ]
}
cr = sq.Circuit(elements)

# set the truncation numbers
cr.set_trunc_nums([25, 25])

# external flux for sweeping over
phi = np.linspace(0, 1,100)

# spectrum of the circuit
n_eig=5
spec = np.zeros((n_eig, len(phi)))

for i in range(len(phi)):
    # set the external flux for the loop
    loop1.set_flux(phi[i])
    
    # diagonlize the circuit
    spec[:, i], _ = cr.diag(n_eig)

# plot the energy spectrum
plt.figure()
for i in range(n_eig):
    plt.plot(phi, spec[i,:] - spec[0,:])
plt.xlabel(r"$\Phi_{ext}/\Phi_0$")
plt.ylabel(r"$f_i-f_0$[GHz]")
plt.show()

# set the external flux back to zero
loop1.set_flux(0)
_, _ = cr.diag(n_eig=2)

# create a range for each mode
phi1 = np.pi*np.linspace(-1,1,100)
phi2 = np.pi*np.linspace(-0.5,1.5,100)

# the ground state
state0 = cr.eig_phase_coord(
    k=0, grid=[phi1, phi2])
    
# the first excited state
state1 = cr.eig_phase_coord(
    k=1, grid=[phi1, phi2])

# plot the ground state
plt.figure()
plt.pcolor(phi1, phi2, 
           np.abs(state0)**2,
           shading='auto')
plt.xlabel(r'$\varphi_1$')
plt.ylabel(r'$\varphi_2$')
plt.show()

# plot the first excited state
plt.figure()
plt.pcolor(phi1, phi2, 
           np.abs(state1)**2,
           shading='auto')
plt.xlabel(r'$\varphi_1$')
plt.ylabel(r'$\varphi_2$')
plt.show()
\end{lstlisting}
and the results of Fig.\ref{fig:loss} by

\begin{lstlisting}[language=python]
# import the libraries
import SQcircuit as sq
import numpy as np
import matplotlib.pyplot as plt

# define the loop of the circuit
loop1 = sq.Loop()

# define the circuit's elements
C = sq.Capacitor(3.6, "GHz", Q=1e6)
L = sq.Inductor(0.46, "GHz", loops=[loop1])
JJ = sq.Junction(10.2, "GHz", cap=C, 
                 A=5e-7, loops=[loop1])

# define the Fluxonium circuit
elements = {(0, 1): [L, JJ]}
cr = sq.Circuit(elements, flux_dist="all")

# set the truncation numbers
cr.set_trunc_nums([100])

# external flux for sweeping over
phi = np.linspace(0, 1, 300)

# T_1 and T_phi
T_1 = np.zeros_like(phi)
T_phi = np.zeros_like(phi)

for i in range(len(phi)):
    # set the external flux for the loop
    loop1.set_flux(phi[i])
    
    # diagonalize the circuit 
    _, _ = cr.diag(n_eig=2)
    
    # get the T_1 for capacitive loss
    T_1[i] = 1/cr.dec_rate(
             dec_type="capacitive",
             states=(1,0))
             
    # get the T_phi for cc noise     
    T_phi[i] = 1/cr.dec_rate(
               dec_type="cc",
               states=(1,0))

# plot the T_1 from the capacitive loss
plt.figure()
plt.semilogy(phi, T_1)
plt.xlabel(r"$\Phi_{ext}/\Phi_0$")
plt.ylabel(r"$s$")
plt.show()

# plot the T_phi from the critical
# current noise
plt.figure()
plt.semilogy(phi, T_phi)
plt.xlabel(r"$\Phi_{ext}/\Phi_0$")
plt.ylabel(r"$s$")
plt.show()

\end{lstlisting}

\bibliographystyle{IEEEtran}
\bibliography{LINQS_qubit_theory}

\end{document}